\documentclass[aps,prb,twocolumn]{revtex4}
\usepackage{epsfig}

\begin{document}

\title{Exact solution for the two-site 
              correlated Kondo-Lattice Model\\
              - a limiting case for a metal}
\author{T.~Hickel} 
\email{hickel@physik.hu-berlin.de}
\author{J.~R\"oseler}
\author{W.~Nolting}
\affiliation{
    Lehrstuhl f\"ur Festk\"orpertheorie, Institut f\"ur Physik,\\
    Humboldt-Universit\"at, 10115 Berlin}
\date{\today}

\begin{abstract}
 The correlated Kondo-lattice model is used to describe the interaction 
 of electrons in a single conduction band with localized magnetic moments 
 as well as their mutual repulsion. It is our intension 
 to provide an analytical exact result for this model by considering 
 a two-site cluster. An equation-of-motion approach as well as the
 spectral theorem is chosen to obtain the complete expression for the
 one-particle Green's function. In a forgoing article the problem
 has been solved for an insulator, the present paper is devoted to the 
 technical more demanding task of a metal.
\end{abstract}
 
\newcommand{\n}{{\hat n}^{ } }
\newcommand{\cd}[1]{c^\dagger_{#1}}
\newcommand{\cn}[1]{c^{ }_{#1}}
\newcommand{\up}{\uparrow}
\newcommand{\dn}{\downarrow}
\newcommand{\ps}{\sigma}
\newcommand{\ms}{{-\sigma}}
\renewcommand{\exp}[1]{\,{\rm exp}\!\left[ #1 \right]}
\newcommand{\ie}{{\rm i}}
\newcommand{\vz}{\epsilon}
\renewcommand{\t}{\tilde T}
\newcommand{\CF}[1]{\big\langle#1\big\rangle}
\newcommand{\GFt}[2]{\big\langle\!\big\langle#1 ;#2
                     \big\rangle\!\big\rangle}
\newcommand{\GF}[2]{\big\langle\!\big\langle#1 ;#2
                    \big\rangle\!\big\rangle_E}
\newcommand{\Gf}[1]{\big\langle\!\big\langle#1
                    \big\rangle\!\big\rangle^{(\vz)}}
\newcommand{\Gff}[1]{\big\langle\!\big\langle#1
                    \big\rangle\!\big\rangle}

\newcommand{\ts}{s}
\newcommand{\os}{{\bar s}}
\newcommand{\EM}{\hat E}
\renewcommand{\a}{a}

\newcommand{\Jh}{\mbox{$\frac{J}{2}$}}
\newcommand{\fh}{\mbox{$\frac{\hbar}{2}$}}
\newcommand{\fgh}{\mbox{$\fh\Jh\,$}}
\newcommand{\Uh}{\mbox{$\frac{U}{2}$}}

\newcommand{\rmI}{{\rm I\!V}}
\newcommand{\rmII}{{\rm I\!I\!I}}
\newcommand{\rmIII}{{\rm I\!I}}
\newcommand{\rmIV}{{\rm I}}

\maketitle
\section{Introduction}
 The magnetism of a large variety of so-called local moment systems, 
 like manganites, gadolinium or europiumchalcogenides, is governed 
 by an interplay of the properties of magnetic moments localized 
 at certain atoms and itinerant conduction electrons propagating 
 through the lattice. If the system of localized spins ${\bf S}_i$
 were treated independently, a certain magnetic order could be explained 
 with the help of a Heisenberg-like exchange coupling 
   \begin{equation}
     \label{eq:Heisenberg}
     {\cal H}_{f\!f} = - \sum_{i,j} J_{ij}\, {\bf S}_i \cdot {\bf S}_j .
   \end{equation}
 In contrast to that the Kondo-lattice model tries to explain 
 magnetic order via an indirect coupling of ${\bf S}_i$ and 
 ${\bf S}_j$ mediated by the spin {\boldmath$\sigma$\unboldmath}
 of the itinerant conduction electrons. The importance of the 
 on-site interaction
   \begin{equation}
     \label{eq:DoubleEx}
     {\cal H}_{sf} = -\frac{J}{\hbar} 
     \sum_i \mbox{\boldmath$\sigma$\unboldmath}_i \cdot {\bf S}_i 
   \end{equation}
 was first emphasized by Zener \cite{Zen51a}.

 As far as the conduction electrons are concerned they might
 be described by their kinetic energy
   \begin{equation}
     \label{eq:Hopping1}
     {\cal H}_s = \sum_{i,j} \sum_\ps T_{ij} \cd{i\ps} \cn{j\ps} .
   \end{equation}
 only. Here, $\cn{j\ps}$ is a fermionic annihilation operator in
 second quantization. However, several authors \cite{KLM+U} 
 pointed out that at least for manganites with its relatively 
 narrow $e_g$-bands an {\it ansatz} that neglects any electron-electron 
 interaction is an insufficient treatment. This is because $U$ is
 probably the largest energy scale in the problem. 

 Another indication of this insufficiency is the exact solution
 of the zero-bandwidth limit (\( T_{ij}= T_0 \delta_{ij} \), 
 also called {\it atomic limit}) of the Kondo-lattice model 
 \( {\cal H}_s + {\cal H}_{sf} \).  
 In this limit the one-particle Green's function has 
 an excitation spectrum of four energy poles \cite{NoM84}
 \begin{equation}
   \label{eq:energies_al}
   \begin{array}{rclrcl}
    \varepsilon_1 &=& T_0 - \fh J S, \quad &
    \varepsilon_2 &=& T_0 + \fh J (S+1), \\
    \varepsilon_3 &=& T_0 - \fh J (S+1), \quad &
    \varepsilon_4 &=& T_0 + \fh J S,
   \end{array}
 \end{equation}
 where $S$ is defined by \( {\bf S}_i^2 = \hbar^2 S (S+1) \).
 If the ``test'' electron enters an empty lattice site
 the energies $\varepsilon_1$ or $\varepsilon_2$ are necessary.
 The other two energies occur if the entered lattice site is 
 already occupied. Apparently $\varepsilon_3$ has the lowest 
 energy for $J>0$. Because of the involved Coulomb repulsion 
 this is not understandable.

 The simplest way to include electron-electron interaction
 in the Hamiltonian is an on-site Hubbard interaction
   \begin{equation}
     \label{eq:Coulomb}
     {\cal H}_{ss} = \frac{U}{2} \sum_i \sum_\ps \n_{i\ps} \n_{i\ms}
               = U \sum_i \n_{i\up} \n_{i\dn} . 
   \end{equation}
 It has been shown \cite{KLM+U} that this has an significant 
 effect on the spectral function of the Kondo-lattice model.
 For the {\it atomic limit} calculation it implies that the
 energies $\varepsilon_3$ and $\varepsilon_4$ are shifted by
 $U$ and $\varepsilon_1$ becomes lowest. 

 The combination of all these possible effects in one Hamiltonian
   \begin{equation}
     \label{eq:CKLM}
     {\cal H} = {\cal H}_s + {\cal H}_{ss} + {\cal H}_{sf} + {\cal H}_{f\!f},
   \end{equation}
 is in this paper called {\bf correlated Kondo lattice model} (CKLM). 
 Due to the fact that there is no analytical solution of the CKLM available,
 it is essential to obtain exact statements from limiting cases. 

 In a foregoing article \cite{HRN01} we explained, why especially
 the generalization of the {\it atomic limit} result to a two-site 
 cluster is important for a better understanding of the CKLM. 
 There, we investigated the Hamiltonian
 \( {\cal H}_s + {\cal H}_{sf} \) with an assumed hopping integral
 of the form
 \[ T_{ij} = \left\{ 
               \begin{array}{ll}
                 \t \quad& i \neq j \,\mbox{and $i,j$ in same cluster,} \\
                T_0 \quad& i=j \,, \\
                  0 \quad& \mbox{$i,j$ in different clusters.}
               \end{array} 
               \right. 
 \]
 The part \( {\cal H}_{f\!f} \) was omitted to gain the chance for
 a connection of the cluster to the lattice. This is possible if
 expectation values of spin operators, like $\CF{S_1^z}$ are not
 uniquely determined by the properties of the cluster. Furthermore,
 a Hamiltonian without \( {\cal H}_{f\!f} \) is much easier to handle.
 The part \( {\cal H}_{ss} \), on the other hand, was redundant
 since the limit of an empty conduction band was considered.
 For this situation, which is related to an insulator, we were
 able to give the complete analytical expression for the one-particle
 Green's function as obtained by a solution of the equations of motion.

 With the present paper we try to obtain similar results for metals.
 Since the occupation of the conduction band is now arbitrary,
 the mathematical systems that need to be solved are about one
 order of magnitude larger than in the case of an insulator.
 Additionally, it is necessary to take the Hubbard interaction
 into account. Therefore, we consider the Hamiltonian
 \begin{eqnarray}
   \bar {\cal H} &=& 
     \t \sum_{\ps=\up,\dn} 
        \left( \cd{1\ps} \cn{2\ps} + \cd{2\ps} \cn{1\ps} \right) 
    +  \sum_{\alpha=1}^2 \sum_{\ps=\up,\dn} 
        \frac{U}{2} \n_{\alpha\ps} \n_{\alpha\ms}  \nonumber \\
   &+& \sum_{\alpha=1}^2 \sum_{\ps=\up,\dn} \Bigg\{
                  T_0 \n_{\alpha\ps}   
                - \frac{J}{2} 
                  \left(z_\ps S^z_{\alpha} \n_{\alpha\ps} 
                      + S^\ps_{\alpha} \cd{\alpha\ms} \cn{\alpha\ps}  
                  \right)
               \Bigg\} ,\nonumber \\[-1ex]
      \label{eq:Cluster-Ham}
 \end{eqnarray}
 where \( S^\up \equiv S^+, S^\dn \equiv S^- \) and 
 \( z_\up \equiv +1,  z_\dn \equiv -1 \). For simplicity
 we restrict ourself to $S=1/2$. 
 
 Despite of the larger effort, we use the same techniques already
 proposed for the insulator. What has been explained on the more
 concise problem there, can now be applied directly. It will not be 
 possible to provide every detail of the calculations here. A 
 thorough documentation of all necessary transformations can be
 found elsewhere \cite{Hic01}.
 
\section{Solution for the equations of motion}
\label{Solution for the equations of motion}
 To describe the properties of the electronic partial system 
 of the cluster we are especially interested in the one-electron
 retarded Green's function \( \GF{\cn{i\ps}}{\cd{j\ps}} \). 
 This function is closely related to the spectral density 
 and contains all information about possible single-particle 
 excitations. 

 The method of choice is an equation-of-motion technique. 
 Compared to a direct determination of the Lehmann 
 representation it has two advantages. First, an equation 
 of motion 
 \begin{equation}
     \label{eq:eom}
      E \GF{\hat A}{\hat B}
         = \hbar \CF { [ \hat A, \hat B ]_+ }
         + \GF{ [ \hat A, \bar {\cal H} ]_-}{\hat B} .       
 \end{equation}
 contains as an inhomogeneity a correlation function 
 \( \CF { [ \hat A, \hat B ]_+ } \). This will directly appear
 in the numerator of the Green's function, without being forced
 to actually calculate the expectation value.
 Therefore, we are able to discuss the dependence of the
 spectral weights on parameters as the averaged band occupation
 or the averaged $z$-component of the localized spins.  
 Secondly, each equation of motion contains higher 
 Green's functions \( \GF{ [ \hat A, \bar {\cal H} ]_-}{\hat B} \).
 Hence, by using intelligent techniques we do not only get 
 \( \GF{\hat A}{\hat B} \) but a complete set of Green's 
 function with much more information. We will make use of this fact
 in the next section.

 Since we are dealing with a finite size system, the hierarchy
 of equations of motion necessarily has to terminate. The complete
 set of 1040 equations can be formulated as a single matrix 
 equation. By unitary transformations this matrix needs to be 
 reduced to blocks of a size which can be solved directly. 
 A first step in this direction is the implementation of the 
 site symmetry. Therefore, we define combined Green's functions 
 \begin{eqnarray}
   \label{eq:comb1}
   \Gf{\hat C_\ts \hat D_\os \cn{\ts\ps}}
   &=& \,\,\, \GF{\hat C_\ts \hat D_\os \cn{\ts\ps}}{\cd{\ts\ps}} \nonumber\\
   &+& \vz    \GF{\hat C_\os \hat D_\ts \cn{\os\ps}}{\cd{\ts\ps}} ,
 \end{eqnarray}
 where the site index $\os$ represents the opposite of site $\ts$
 ($\ts=1 \Rightarrow \os=2; \ts=2 \Rightarrow \os=1$), $\hat C_\ts$ and 
 $\hat D_\os$ are arbitrary products of spin- and Fermi-operators. 
 The two partial systems for \( \vz = + 1 \) and \( \vz = -1 \) can be 
 solved separately and simultaneously (they only differ in the sign of
 $\t$). 

 Even if additionally the mirror symmetry with respect to the
 spin quantization axis is used, one is still left with the large 
 number of Green's function listed below:
  {\setlength{\arraycolsep}{1pt}
  \[\begin{array}{rclrcl}
   G_{\rm A}^{(\vz)}\! &=&
   \Gf{\hat S \cn{\ts\ps}}
   & K_{\rm A}^{(\vz)}\! &= 
         & \CF{\hat S}  \\
   G_{\rm B}^{(\vz)}\! &=&
   \Gf{\hat S \n_{\ts\ms} \cn{\ts\ps}}
   & K_{\rm B}^{(\vz)}\! &= 
         & \CF{\hat S \n_{\ts\ms}}   \\
   G_{\rm C}^{(\vz)}\! &=&
   \Gf{\hat S \n_{\os\ms} \cn{\ts\ps}}
   & K_{\rm C}^{(\vz)}\! &= 
         & \CF{\hat S \n_{\os\ms}}  \\
   G_{\rm D}^{(\vz)}\! &=&
   \vz \Gf{\hat S \cd{\ts\ms} \cn{\os\ms} \cn{\ts\ps}}
   & K_{\rm D}^{(\vz)}\! &= 
         & \vz \CF{\hat S \cd{\ts\ms} \cn{\os\ms}}  \\
   G_{\rm F}^{(\vz)}\! &=&
   \vz \Gf{\hat S \cd{\os\ms} \cn{\ts\ms} \cn{\ts\ps}}
   & K_{\rm F}^{(\vz)}\! &= 
         & \vz \CF{\hat S \cd{\os\ms} \cn{\ts\ms}}   \\
   G_{\rm R}^{(\vz)}\! &=&
   \Gf{\hat S \n_{\os\ps} \cn{\ts\ps}}
   & K_{\rm R}^{(\vz)}\! &= 
         & \CF{\hat S \n_{\os\ps}}
   \\&&&&-&
        \vz\CF{\hat S \cd{\ts\ps} \cn{\os\ps}} \\
   G_{\rm G}^{(\vz)}\! &=&
   \Gf{\hat S \n_{\os\ms} \n_{\ts\ms} \cn{\ts\ps}}
   & K_{\rm G}^{(\vz)}\! &= 
         & \CF{\hat S \n_{\os\ms} \n_{\ts\ms}}  \\
   G_{\rm H}^{(\vz)}\! &=&
   \vz \Gf{\hat S \n_{\os\ps} \cd{\ts\ms} \cn{\os\ms} \cn{\ts\ps}}
   & K_{\rm H}^{(\vz)}\! &= 
         & \vz \CF{\hat S \n_{\os\ps} \cd{\ts\ms} \cn{\os\ms}} 
   \\&&&&-&
         \CF{\hat S \cd{\os\ms} \cn{\ts\ms} \cd{\ts\ps} \cn{\os\ps}} \\
   G_{\rm J}^{(\vz)}\! &=&
   \vz \Gf{\hat S \n_{\os\ps} \cd{\os\ms} \cn{\ts\ms} \cn{\ts\ps}}
   & K_{\rm J}^{(\vz)}\! &= 
         & \vz \CF{\hat S \n_{\os\ps} \cd{\os\ms} \cn{\ts\ms}}
   \\&&&&-&
         \CF{\hat S \cd{\ts\ms} \cn{\os\ms} \cd{\ts\ps} \cn{\os\ps}} \\
   G_{\rm K}^{(\vz)}\! &=&
   \Gf{\hat S \n_{\os\ps} \n_{\ts\ms} \cn{\ts\ps}}
   & K_{\rm K}^{(\vz)}\! &= 
         & \CF{\hat S \n_{\os\ps} \n_{\ts\ms}}
   \\&&&&-&
       \vz\CF{\hat S \n_{\os\ms} \cd{\ts\ps} \cn{\os\ps}} \\
   G_{\rm L}^{(\vz)}\! &=&
   \Gf{\hat S \n_{\os\ps} \n_{\os\ms} \cn{\ts\ps}}
   & K_{\rm L}^{(\vz)}\! &= 
         & \CF{\hat S \n_{\os\ps} \n_{\os\ms}}
   \\&&&&-&
       \vz\CF{\hat S \n_{\ts\ms} \cd{\ts\ps} \cn{\os\ps}} \\
   G_{\rm M}^{(\vz)}\! &=&
   \Gf{\hat S \n_{\os\ms} \n_{\os\ps} \n_{\ts\ms} \cn{\ts\ps}}
   & K_{\rm M}^{(\vz)}\! &= 
         & \CF{\hat S \n_{\os\ms} \n_{\os\ps} \n_{\ts\ms}}
   \\&&&&-&
       \vz\CF{\hat S \n_{\ts\ms} \n_{\os\ms} \cd{\ts\ps} \cn{\os\ps}} \\
  \end{array}\]}\\
 Here, $\hat S$ stands for all possible products of spin operators.
 If, as assumed, $S=1/2$ there are 10 such possibilities. 
 The right column gives for each Green's function the corresponding
 correlation function emerging in its equation of motion. 
 It is useful to group these functions into four {\it density classes}, determined
 by the number of fermionic creation operators in its active part:
 \begin{eqnarray*}
   G_\rmIV &:& G_{\rm A}, \\ 
   G_\rmIII &:& G_{\rm B}, G_{\rm C}, G_{\rm D}, G_{\rm F}, G_{\rm R}, \\
   G_\rmII &:& G_{\rm G}, G_{\rm H}, G_{\rm J}, G_{\rm K}, G_{\rm L}, \\
   G_\rmI &:& G_{\rm M} .
 \end{eqnarray*}
 The Green's functions of density class $\rmI$ vanish whenever the
 number of participating electrons is smaller than 3. The ones in 
 density class $\rmII$ vanish if we have less than 2 electrons etc.. 
 Using this notation the matrix equation of all Green's functions has
 the block structure
 {\setlength{\arraycolsep}{0pt} 
 \begin{equation} 
   \label{eq:matrix}
   \left(\begin{array}{cccc} 
      \mbox{\fbox{$M_\rmIV$}} & M_{\rmIV,\rmIII} & 0 & 0 \\
      0 & \mbox{\fbox{\( \,\frac{ }{ } M_\rmIII \,\)}} & M_{\rmIII,\rmII} & 0 \\
      0 & 0 & \mbox{\fbox{\( \,\frac{ }{ } M_\rmII \,\)}} & \,M_{\rmII,\rmI}  \\
      0 & 0 & 0 & \mbox{\fbox{$M_\rmI$}} 
   \end{array}\right)
   \left( \begin{array}{ccc}  G_\rmIV  \\[0.5ex] \hline \\[-4ex]
                              G_\rmIII \\[0.5ex] \hline \\[-4ex]
                              G_\rmII  \\[0.5ex] \hline \\[-4ex]
                              G_\rmI   
          \end{array} \right)
  =\left( \begin{array}{ccc}  K_\rmIV  \\[0.5ex] \hline \\[-4ex]
                              K_\rmIII \\[0.5ex] \hline \\[-4ex]
                              K_\rmII  \\[0.5ex] \hline \\[-4ex]
                              K_\rmI   
          \end{array} \right) ,
 \end{equation}}\\
 where the right hand side represents the column of all inhomogeneities
 appearing in every equation of motion. 

 Since the matrix has a triangular shape, the remaining task is to find the
 inverse of the blocks \( M_\rmIV, M_\rmIII, M_\rmII \) and \( M_\rmI \).
 It is therefore possible to stay in certain density classes, 
 which justifies the grouping. Additionally,
 we can benefit from particle-hole symmetry. It implies that the density
 classes $\rmIV$ and $\rmI$ as well as the classes $\rmIII$ and $\rmII$ 
 can be treated in a similar way, respectively.

 The density class $\rmIV$ consists of the 10 Green's functions with 
 only one fermionic operator but with several spin operators. If one could
 assume that all Green's functions of the other density classes vanish,
 one just had to invert the matrix $M_\rmIV$ to get a solution. 
 The assumption is fulfilled if and only if one describes an insulator 
 with a single excess electron in a otherwise empty conduction band. 
 The foregoing article \cite{HRN01} was devoted to these materials. 
 In the case of arbitrary band occupations we have to solve the equation
 \( M_\rmIV \cdot G_\rmIV = K_\rmIV - M_{\rmIV,\rmIII} G_\rmIII \).
 Since only the right hand side is modified we can use the same
 rules for the combination of spin operators in the Green's functions
 as in the case of an insulator to simplify $M_\rmIV$.
 
 Matrix $M_\rmIII$ is also simplified by an intelligent combination
 of Green's functions. As far as the spin operators are
 concerned we use again the rules proposed in the treatment of 
 an insulator. To get an idea about the behavior of the 
 fermionic operators we investigated another limiting case: If in 
 $\bar {\cal H}$ the parameter $J$ is set to $0$ we have a system 
 without any spin operator, namely a Hubbard cluster. Its solution is 
 sketched in appendix \ref{The Hubbard cluster}. The rules for a 
 formation of combined Green's functions used there can then be
 applied to equation (\ref{eq:matrix}).

 This technique to combine Green's functions by a successive 
 treatment of the spin operators and the fermionic operators
 respectively works, but is rather lengthy and fault-prone.
 When doing this job we learned that a simple text editor
 can be awesomely helpful. Already when generating the 1040 
 equations of motion a {\tt cut-and-paste} mechanism can be used 
 effectively, since the same commutators
 \( [ \cn{\ts\ps}, \bar {\cal H} ]_-, 
    [ S^z_\ts, \bar {\cal H} ]_- \) and
 \( [ S^\ps_\ts, \bar {\cal H} ]_- \)
 have to be implemented several times. For combinations the 
 {\tt search-and-replace} function is quite helpful. 

 Appendix \ref{Combinations of Green's functions} gives 
 detailed instructions which Green's functions need to be
 summed and subtracted to achieve a simpler structure
 of $M_\rmIII$. If implemented one obtains matrix 
 blocks of size $3 \times 3$. 
 There are four types of these matrices:
 \begin{eqnarray}
     \label{eq:Matrix0}
     M_0^{(\mu_1 \mu_2 \mu_3 \mu_4)} (\EM) &=& 
     \left(
       \begin{array}{ccc}
          \EM+2\mu_4 \a    & \mu_1 \t  &  0 \\
          \mu_1 \t  & \EM       & \mu_2 2\a \\
          0         & \mu_2 2\a & \EM+\mu_3 \t    
       \end{array}
     \right) , \\
     \label{eq:Matrix1}
     M_1^{(\mu_2 \mu_5)} (\EM) &=& 
     \left(
       \begin{array}{ccc}
          \EM+ \mu_5 U  & 0 &  0 \\
          0 & \EM       & \mu_2 2\a \\
          0 & \mu_2 2\a & \EM    
       \end{array}
     \right) , \\
     \label{eq:Matrix2}
     M_2^{(\mu_1 \mu_2 \mu_5)} (\EM) &=& 
     \left(
       \begin{array}{ccc}
          \EM+ \mu_5 U     & \mu_1 2\t &  0 \\
          \mu_1 2\t & \EM       & \mu_2 2\a \\
          0         & \mu_2 2\a & \EM    
       \end{array}
     \right) , \\
     M_3^{(\mu_6)} (\EM) &=& 
     \left(
       \begin{array}{ccc}
          \EM+\mu_6 U  &  2\t &  0 \\
          2\t  & \EM   & 6\mu_6 \a \\
          0  & 2\mu_6 \a   & \EM+4\mu_6\a    
       \end{array}
     \right) .\quad
 \end{eqnarray}
 Here, we used the abbreviation \( \a = \frac{\hbar}{2}\frac{J}{2} \)
 and \( \mu_1, \ldots, \mu_6 \) are sign parameters. 
 $M_0$, which already played a role in the insulator problem \cite{HRN01}, 
 has the eigenvalues
 \begin{eqnarray*}
  \EM_{01}^{(\mu_3 \mu_4)} &=& \mu_3 \t +2\mu_4 \a , \\
  \EM_{02}^{(\mu_3 \mu_4)} &=& -\sqrt{ 4\a^2 - 2\mu_3\mu_4\a\t + \t^2} , \\
  \EM_{03}^{(\mu_3 \mu_4)} &=& +\sqrt{ 4\a^2 - 2\mu_3\mu_4\a\t + \t^2} .
 \end{eqnarray*}
 The other matrices are connected to non-vanishing 
 electron-densities as can be seen at the appearance of $U$.
 The structure of $M_1$ is simple, and its eigenvalues are
 \begin{equation}
   \label{eq:EW1}
   \EM_{11}^{(\mu_5)} = \mu_5 U, \qquad
   \EM_{12} = 2\a, \qquad \EM_{13} = -2\a.
 \end{equation}
 In contrast to that the determinantal polynomial of $M_2$
 \begin{equation}
    \label{eq:cP_Matrix2}
    0 \stackrel{\rm !}{=} 
    \EM^3 + \mu_5 \EM^2 U - 4\EM \left(\a^2+\t^2 \right) - 4\a^2 \mu_5 U
 \end{equation}
 can only be factorized with Cardan's formulas. The three eigenvalues 
 of this cubic equation are
 \begin{eqnarray}
   \label{eq:EW21}
   \EM_{21}^{(\mu_5)} &=& \mu_5 \frac{U}{3} - \frac{2}{3} \Re{\rm e\,} C_2 ,\\
   \label{eq:EW22}
   \EM_{22}^{(\mu_5)} &=& \mu_5 \frac{U}{3} + \frac{1}{3} 
             \left[ \Re{\rm e\,} C_2 + \sqrt{3}\, \Im{\rm m\,} C_2 \right] \\
   \label{eq:EW23} \mbox{and~~}
   \EM_{23}^{(\mu_5)} &=& \mu_5 \frac{U}{3} + \frac{1}{3} 
             \left[ \Re{\rm e\,} C_2 - \sqrt{3}\, \Im{\rm m\,} C_2 \right] ,\\
   \mbox{where~~}
   C_2 &=& \sqrt[3]{\mu_5 A_2 + \ie \sqrt{B_2^3-A_2^2}} \nonumber \\
   \mbox{and~~}
   A_2 &=&  36 \a^2 U - 18 \t^2 U - U^3  \nonumber \\
   B_2 &=&  12 \a^2 + 12 \t^2 + U^2  .    \nonumber
 \end{eqnarray}
 Similarly, we have for $M_3$ the cubic equation
 \begin{eqnarray}
    \label{eq:cP_Matrix3} \nonumber
    0 &\stackrel{\rm !}{=}&
    \EM^3 + \mu_6 \EM^2 (4\a+U) - 4\EM \left(3\a^2+\t^2 -\a U \right) 
    \\&&  - \mu_6 4\a \left( 4\t^2 + 3\a U \right) ,
 \end{eqnarray}
 with the three solutions
 \begin{eqnarray}
    \EM_{31}^{(\mu_6)} &=&  \frac{\mu_6}{3}( U+4\a) - \frac{2}{3} \Re{\rm e\,} C_3 ,\\
    \EM_{32}^{(\mu_6)} &=&  \frac{\mu_6}{3}( U+4\a) + \frac{1}{3} 
              \left[ \Re{\rm e\,} C_3 + \sqrt{3} \Im{\rm m\,} C_3
              \right] \\
    \mbox{and~~}
    \EM_{33}^{(\mu_6)} &=&  \frac{\mu_6}{3}( U+4\a) + \frac{1}{3} 
              \left[ \Re{\rm e\,} C_3 - \sqrt{3} \Im{\rm m\,} C_3
              \right] , \quad
 \end{eqnarray}
 where \( C_3 = \sqrt[3]{\mu_6 A_3 + \ie \sqrt{B_3^3 - A_3^2}} , \)
 \begin{eqnarray}
     A_3 &=& -280 \a^3 + 132 \a^2  U - 18 \t^2  U 
             - U^3 + 6\a(24\t^2+U^2) \nonumber \\
     B_3 &=&  52\a^2 + 12\t^2 - 4\a U +  U^2  .\nonumber
 \end{eqnarray}

 If the eigenvalue problem of the cluster is studied and especially 
 the Hilbert subspace of two electrons in the conduction band is 
 considered, one is confronted with exactly the same two cubic equations 
 (\ref{eq:cP_Matrix2}) and (\ref{eq:cP_Matrix3}). This underlines the 
 close relationship of the eigenvalue problem and the equation-of-motion
 technique.

 To determine the inverse of $M_x, x=0,\ldots,3$ it is sufficient
 to know its eigenvectors. As explained more detailed in the 
 treatment of the insulator \cite{HRN01}, basic algebra then
 immediately allows to give an expression for $M_x^{-1}$:
 \begin{equation}
   \label{eq:inverse}
   M_x^{-1}[ij] = \sum_{k=1}^3 \frac{m_{xk}(i,j)}{\EM + \EM_{xk}},
 \end{equation}
 with well defined coefficients $m_{xk}(i,j)$. This structure is
 particularly suitable because we intend to get the analytic expression 
 for the one-particle Green's function as
 \begin{equation}
   \label{eq:standardform}
   G(E) = \GFt{\hat A}{\hat B}_{E+\ie 0^+}
        = \sum_{k=1}^p \frac{\hbar \alpha_k}{E-E_k+\ie 0^+},
 \end{equation}
 with $p$ linear energy poles.
 Since we succeeded to find combined Green's functions which are
 connected in $3 \times 3$ systems, an inversion of the matrices
 like in (\ref{eq:inverse}) will ensure that every combined 
 Green's function has the form (\ref{eq:standardform}).
 To return to the original Green's functions one just has to undo 
 the combinations. That happens by summations and subtractions of 
 the combined expressions and will therefore maintain the form 
 (\ref{eq:standardform}).

 We mentioned above that the calculations can be done within certain 
 density classes. To be more precise, to solve equation (\ref{eq:matrix})
 the following four matrix equations have to be considered: 
 \begin{equation}
    \label{eq:steps}
    \begin{array}{rllcl}
     \rmI &:\qquad& M_\rmI \cdot G_\rmI &=& K_\rmI \\
     \rmII &:& M_\rmII \cdot G_\rmII &=& K_\rmII - M_{\rmII,\rmI} G_\rmI \\
     \rmIII &:& M_\rmIII \cdot G_\rmIII &=& K_\rmIII - M_{\rmIII,\rmII} G_\rmII \\
     \rmIV &:& M_\rmIV \cdot G_\rmIV &=& K_\rmIV - M_{\rmIV,\rmIII} G_\rmIII 
    \end{array}
 \end{equation}
 One has to obey this order, because apart from system $\rmI$ we
 always have an input of a class with higher electron densities.
 On the one hand, the matrix multiplication on the right side implies
 an additional summation of already calculated Green's functions. 
 On the other hand, a partial fraction expansion is necessary
 to return to the form (\ref{eq:standardform}).
 \begin{eqnarray}
   \frac{K_\rmII-\frac{K_\rmI}{E-E_\rmI}}{E-E_\rmII}
  = \frac{K_\rmII}{E-E_\rmII} - \frac{K_\rmI}{(E-E_\rmII)(E-E_\rmI)} \qquad \\
  = \frac{K_\rmII}{E-E_\rmII} + \frac{K_\rmI}{E_\rmI-E_\rmII} 
               \left(\frac{1}{E-E_\rmII} - \frac{1}{E-E_\rmI} \right) \nonumber
 \end{eqnarray}
 Occasionally, it happens that $E_\rmII = E_\rmI$, leading to quadratic
 energy poles. However, after performing all necessary summations
 the spectral weight of these poles will always vanish. 

 Taking all this into account, one obtains after a
 straightforward calculation expressions for the 
 Green's functions of density class $\rmIV$. If furthermore
 all combinations are undone, the desired analytical exact 
 result for the one-electron Green's function 
 \( \GF{\cn{\ts\ps}}{\cd{\ts\ps}} \) can be given. 
 It consists of 102 energy poles.

 Six of them, namely the energies
 \begin{eqnarray}
   \label{eq:poles01}
   E &=& T_0 + \a - \EM_{0k}^{(++)} \quad \mbox{and} \\
   \label{eq:poles02}
   E &=& T_0 + \a - \EM_{0k}^{(-+)} \quad \mbox{for~} k=1,2,3,
 \end{eqnarray}
 are a result of the inversion of $M_\rmIV$.
 They are just a subset of the energies 
 \begin{eqnarray}
   \label{eq:poles11}
   E &=& T_0 -\a - \EM_{1l}^{(-)} - \EM_{0k}^{(\pm-)} ,\\
   \label{eq:poles12}
   E &=& T_0 -\a - \EM_{2l}^{(-)} - \EM_{0k}^{(\pm-)} \quad\mbox{and}\\
   \label{eq:poles13}
   E &=& T_0 -\a - \EM_{3l}^{(-)} - \EM_{0k}^{(\pm-)} \quad
   \mbox{for~}k,l=1,2,3
 \end{eqnarray}
 which occur, if $M_\rmIII$ is inverted.
 The structure of these energy poles is quite clear. 
 The Green's functions of density class $\rmIII$ include
 excitation processes from a one-electron state to a two-electron
 state. The energies of the first are mediated by the
 eigenvalues of $M_0$, the energies of the latter are 
 connected to the matrices $M_1, M_2$ and $M_3$. $T_0$ is
 just a trivial energy shift. $-\a$ actually belongs into
 the definition of $M_0$. The super-index ``$\pm$'' in $\EM_{0k}$
 illustrates that always both possible signs of $\t$ need to
 be considered, as mentioned subsequently to (\ref{eq:comb1}). 
 Hence, $M_\rmIII$ leads to $3 \times 3 \times 3 \times 2 = 54$
 energy poles.

 It has already been pointed out, that $M_\rmII$ can be 
 treated similarly to $M_\rmIII$ according to the particle-hole
 symmetry of the system. This symmetry can be investigated more
 thoroughly when the Hamiltonian is reformulated in terms
 of hole-creation and -annihilation operators, 
 \( b^\dagger_{i-\ps} = z_\ps \cn{i\ps} \). It leads to the
 result that with an energy pole \( E_k = T_0 + f(\t, J, U) \)
 of the one-particle Green's function, 
 \( \bar E_k = U + T_0 - f(-\t, J, U) \)
 is an energy pole, too. Applied to (\ref{eq:poles11})-
 (\ref{eq:poles13}) one gets all the energies related to the
 matrices $M_\rmII$ and $M_\rmI$. Since
 \begin{equation}
   \label{eq:degeneracy}
     T_0 -\a - \EM_{11}^{(-)} - \EM_{0k}^{(\pm-)} 
   = T_0 + U + \a + \EM_{1l}^{(-)} + \EM_{0k'}^{(\mp-)}, 
 \end{equation}
 we have six degenerate energies, leading to
 $p = 2 \times 54 -6 = 102$ energy poles of the one-particle
 Green's function \( \GF{\cn{i\ps}}{\cd{i\ps}} \) and 
 all higher Green's functions.

\begin{figure}[ht]    
    \epsfig{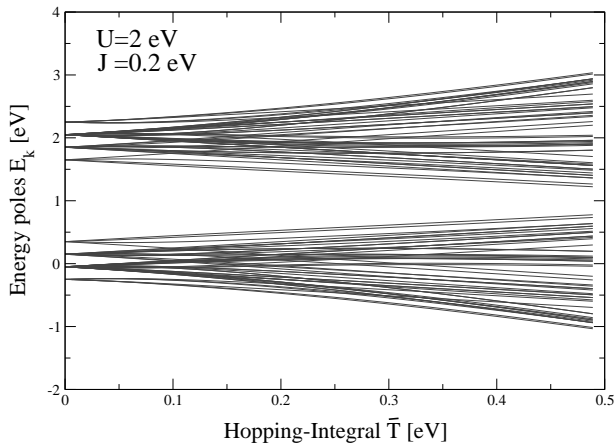}
    \caption[Dependence of energy poles on small T]
    {Dependence of every energy poles of the one-particle Green's
     function on the hopping integral $\t$. The fixed model parameters
     are: $T_0= 0 {\rm\, eV}, U = 2 {\rm\, eV}$ and $J = 0.2 {\rm\, eV}$.
     \label{fg:T-dep1} }
 \end{figure}

\begin{figure}[ht]    
    \epsfig{file=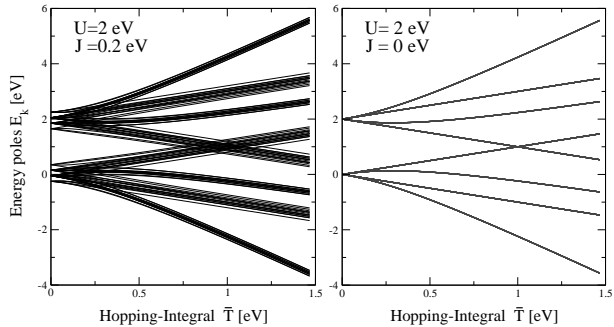, width=0.45\textwidth}
    \caption[Dependence of energy poles on large T]
    {Dependence of the energy poles on the hopping integral $\t$. 
     At the left the behaviour of the CKLM is shown, for comparison
     the Hubbard cluster ($J=0$) is investigated at the right for the 
     same parameter range.
     \label{fg:T-dep2} }
 \end{figure}

\begin{figure}[ht]    
    \epsfig{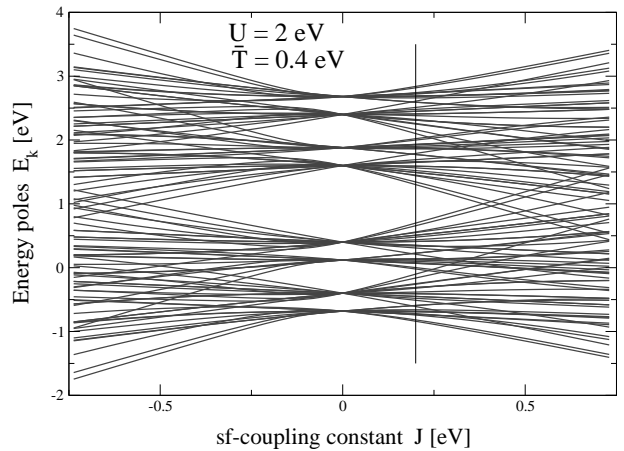}
    \caption[Dependence of energy poles on J]
    {Dependence of the energy poles on the sf-coupling constant $J$.
     The fixed model parameters are: 
     $T_0= 0 {\rm\, eV}, U = 2 {\rm\, eV}$ and $\t = 0.4 {\rm\, eV}$.
     The vertical line marks the $J$ value of figure \ref{fg:T-dep1}.
     \label{fg:J-dep1} }
 \end{figure}

 All these excitation energies are functions of the model 
 parameter $\t, U$ and $J$. Since the cluster is a generalization
 of the {\it atomic limit}, the $\t$-dependence is of special
 interest (see figure \ref{fg:T-dep1}). 
 Because of the large number of energy lines, the formation
 of two Hubbard bands can clearly be seen. Although they are
 well separated at small $\t$, there seems to be an overlap of the
 bands above $0.5 {\rm ~eV}$. When going to higher values
 of $\t$ a separation of the bands and a formation of bundles 
 occurs. As can be seen when the two parts of figure \ref{fg:T-dep2} 
 are compared
 each of these bundles results from the $J$-splitting of an energy pole 
 obtained for the two-site Hubbard cluster (appendix \ref{The Hubbard cluster}).
 This splitting can best be shown, if, as in figure \ref{fg:J-dep1}, the
 $J$-dependence of the energy poles is studied. The picture suggests a 
 symmetry with respect to the sign of $J$, but this is not completely given. 

 Coming back to figure \ref{fg:T-dep1} another striking fact needs to
 be mentioned. It concerns the limit $\t \to 0$, which is identically to 
 the situation of the {\it atomic limit}. As noted in (\ref{eq:energies_al})
 we expect in this limit a reduction to four energy poles. In contrast
 to that we obtain twice as many energy lines for $\t=0$ in figure 
 \ref{fg:T-dep1}. Obviously, the energies alone are insufficient to
 describe excitation processes. Remarks on the corresponding spectral
 weights are necessary. 

 The routine we suggested for the solution of the equations of
 motion inevitably gives for each Green's function these spectral 
 weights, too. The fact that we don't give the complete analytical 
 solution here, has two main reasons. 
 First of all, the routine consists of quite a lot of complicated 
 summations: The combined Green's function for instance are just sums. 
 At least one summation is necessary to get the inverse of a single 
 matrix as given by equation (\ref{eq:inverse}). Their
 multiplication with a column of correlation functions as well as
 the matrix multiplications in each of the lines of (\ref{eq:steps})
 imply further sums. At the end partial fraction expansions have to be
 performed and the combinations of Green's functions needs to be 
 undone. Even so, all this is simple arithmetics it leads to really
 lengthy expressions for the spectral weights. 
 Secondly, the spectral weights of each Green's function depend
 in principle on all correlation functions that appear in inhomogeneities
 of the same or higher density classes. For \( \GF{\cn{1\up}}{\cd{1\up}} \)
 this means a dependence on 1040 correlation functions (1/3 is
 zero by definition). All of them need to be specified. Only
 afterwards the result for the spectral weights is really meaningful. 
 
\section{Treatment of the correlation functions}

 Already when dealing with the limiting case of an empty conduction
 band \cite{HRN01} one has to develop concepts for the determination of
 the emerging spin correlation functions. On the one hand, 
 a simpler Hamiltonian, that describes the spin system and includes 
 lattice effects like anisotropy or molecular fields, can be investigated. 
 Alternatively, the spectral theorem provides information on these
 correlation functions. The latter concept implies that all calculations
 are based on the cluster Hamiltonian only and no connection to the 
 lattice is made. In this paper we are going to use a combination 
 of these two concepts.

 As mentioned before, it is a crucial advantage of our methods that 
 besides \( \GF{\cn{i\ps}}{\cd{i\ps}} \) a complete set of all Green's 
 functions participating in the equations of motion is obtained.
 For an arbitrary Green's function, by construction given in the form 
 (\ref{eq:standardform}), the spectral theorem allows the calculation
 of a corresponding correlation function
  \begin{equation}
    \label{eq:spectraltheorem}
     \CF{\hat{B} \hat{A}} = \sum\limits_{k=1}^p \alpha_k \, f_-(E_k) 
          \quad = \sum\limits_{k=1}^p 
                    \frac{\alpha_k}{ {\rm e}^{\beta (E_k-\mu)} + 1 } \,,
  \end{equation}
 where $\beta = (k_{\rm B}T)^{-1}$. Therefore, we obtain expressions
 for 1040 correlation functions (again 1/3 is zero by definition). 
 Their principle structure is equal for Green's functions of the
 same density class. Table \ref{tab:Korrel} allows a comparison with
 the structure of the corresponding inhomogeneities.
  \begin{table}[ht]
    \[\begin{array}{c|c|c}
        \mbox{density class} & \mbox{inhomogeneity}
       & \mbox{spectral theorem} \\ \hline && \\[-2ex]
        \rmI & \CF{\hat S\, \hat n^3} & \CF{\hat S\, \hat n^4} \\
        \rmII & \CF{\hat S\, \hat n^2} & \CF{\hat S\, \hat n^3} \\
        \rmIII & \CF{\hat S\, \hat n^1} & \CF{\hat S\, \hat n^2} \\
        \rmIV & \CF{\hat S\, \hat n^0} = \CF{\hat S\,} & \CF{\hat S\, \hat n^1} 
    \end{array}\]
    \caption[Table of correlation functions]
    {For Green's functions of every density class the principle structure
     of the correlation functions emerging in the equation of motion 
     (\ref{eq:eom}) and obtained by the spectral theorem 
     (\ref{eq:spectraltheorem}) is given. 
     \label{tab:Korrel}}
  \end{table}

 The application of the spectral theorem to Green's functions of density class
 $\rmI$ results in correlation functions that do not appear in the spectral
 weights we are looking for. Only the application to the $960$ Green's functions 
 of the remaining density classes is helpful. For each of them the spectral
 theorem provides an equation that is linear in the correlation functions, 
 because this linearity is given for the spectral weights $\alpha_k$
 in (\ref{eq:spectraltheorem}).  
 In principle we can therefore write a homogeneous matrix equation for
 the column of correlation functions $K_i$: 
  \begin{equation}
    \label{eq:LGS1}
    \left(\begin{array}{ccc}
           a_{1,1} & \ldots & a_{1,1040} \\
           \vdots  & \ddots & \vdots \\
           a_{960,1} & \ldots & a_{960,1040} 
    \end{array}\right)
    \mbox{\small \(
       \left( \begin{array}{c} K_{1} \\ \vdots \\[-1ex] \vdots \\ 
                               K_{1040} \end{array} \right) \)}
     = \left( \begin{array}{c} 0 \\ \vdots \\ 0 \end{array} \right).
  \end{equation}
 
 Written in this form, equation (\ref{eq:LGS1}) is underdetermined.
 This seems to offer the possibility to use additionally the other concept for
 the correlation functions, that is to make assumptions for certain 
 expectation values or to specify them on the basis of a Hamiltonian 
 that includes lattice effects. To achieve a non-vanishing magnetic order 
 of the system of localized spins it is necessary to specify the
 spin correlation functions, $\CF{\hat S\,}$. Lattice effects can
 be included if these expectation values are based on the Hamiltonian
 ${\cal H}_{f\!f} + {\cal H}_{f}$ with 
  \begin{equation}
    \label{eq:bfield}
    {\cal H}_f = - b \left( S_1^z + \eta S_2^z \right) , \quad
    \eta = \pm 1 ,
  \end{equation}
 as discussed in the forgoing article \cite{HRN01}. 

 Whenever a certain correlation function is specified, the matrix 
 equation (\ref{eq:LGS1}) becomes inhomogeneous:
 {\setlength{\arraycolsep}{0pt}
  \begin{equation}
    \label{eq:LGS2}
    \left(\begin{array}{ccccc}
           a_{1,1} & \ldots & a_{1,x-1}\!\!\! & a_{1,x+1} & \ldots  \\
           \vdots  & & \vdots & \vdots &  \\[-1ex]
           \vdots  & & \vdots & \vdots &  \\
           a_{960,1} & \ldots & a_{960,x-1} & a_{960,x+1} & \ldots  
    \end{array}\right) \!\!
       \mbox{\footnotesize \(
       \left( \begin{array}{c} K_{1} \\ \vdots \\ K_{x-1} \\ K_{x+1}
                            \\ \vdots \\ K_{1040} \end{array} \right) \)}
     = K_{x} \left( \begin{array}{c} a_{1,x} \\ \vdots \\[-1ex] 
                                    \vdots \\ a_{960,x} \end{array} \right).
  \end{equation}
 }\\
 With the inhomogeneity in the equation of motion of 
 \( \GF{\cn{i\ps}}{\cd{i\ps}} \), which is $1$ by definition, at least
 one specification is inevitable, enforcing a nontrivial result for the
 set \( \left\{ K_i \right\}_i \).

 However, the freedom for the specification of certain correlation 
 functions is drastically reduced because of additional information 
 available.  We have already mentioned that 1/3 of the $K_i$'s is 
 zero by definition, which reduces the number of unknown variables
 in (\ref{eq:LGS1}). Even so, in 1/3 of the cases the application
 of the spectral theorem (\ref{eq:spectraltheorem}) does also give 
 zero at the left hand side, this does not lead to a reduction of 
 the number of equations since the right hand side is still a linear
 relation of the variables. Furthermore, we know that certain 
 pairs of correlation functions have to be identical due to 
 symmetrical reasons.  

 Hence, we actually do not have an under-determined problem,
 but rather more equations than variables. The indicator for
 a proper choice of equations is a nonzero result for the
 determinant of the matrix in (\ref{eq:LGS2}). Rules can again
 best been studied, if the limiting case $J=0$, the Hubbard 
 cluster, is considered. 

 For an overdetermined problem the introduction of additional
 specifications for correlation functions is
 of course problematic. They are likely to contradict some 
 other information we have about the system. 
 For the sake of a connection between cluster and lattice
 we want to accept that. The consequences shall be demonstrated
 with an example: If the atomic limit ($\t=0$) is considered, 
 7 nontrivial correlation functions appear in the spectral 
 weights of the one-particle Green's function. Since the 
 spectral theorem provides only 6 equations, $\CF{S^z_1}$
 can be specified on the basis of a Brillouin function for
 the lattice. However, if a non-vanishing value for $\CF{S^z_1}$
 is chosen one has to be aware of the fact that the equality 
 \( \CF{S^+_1 \cd{1\dn}\cn{1\up}}=\CF{S^-_1 \cd{1\up}\cn{1\dn}} \)
 will be broken. The equality itself follows from the 
 assumption that expectation values are real, but can also be derived
 from particle-hole symmetry. 

 If an equation like (\ref{eq:LGS2}) needs to be solved exactly 
 for the correlation functions \( \left\{ K_i \right\}_i \) standard 
 algebraic techniques like Gauss' elimination method can be 
 used. In our work we implemented all these tedious manipulations 
 in a computer program. Its input values are the model parameters
 and assumed specifications for the correlation functions. 
 By setting successively one and only one correlation function to 
 $1.0$, calculating the set of Green's functions (here the 
 analytical exact expressions are used) and applying the
 spectral theorem, the matrix (\ref{eq:LGS1}) is determined. 
 Afterwards the matrix equation (\ref{eq:LGS2}) is solved 
 numerically exact. 

 The last step contains a specific numerical problem.
 Even if a proper set of equations is selected, for large
 ranges for the chemical potential $\mu$ the determinant of 
 the matrix in (\ref{eq:LGS2}) might be rather small. 
 This is connected to the fact that the Fermi distribution 
 function in (\ref{eq:spectraltheorem}) is almost a step 
 function for low temperatures. The effect can already be
 studied when looking at the {\it atomic limit} for the Hubbard 
 model
 \begin{equation}
   \label{eq:numexHam}
   {\cal H} = T_0 \sum_\ps \n_\ps + \frac{U}{2} \sum_\ps \n_\ps \n_\ms .
 \end{equation}
 After the application of (\ref{eq:spectraltheorem}) the 
 two equations ($\ps=\up,\dn$) 
 \begin{equation}
   \label{eq:numexCF}
   \CF{\n_\ps} = \left( 1-\CF{\n_\ms} \right) f_-(T_0) 
               +  \CF{\n_\ms} f_-(T_0+U)     
 \end{equation}
 define the correlation functions. Whenever $\mu$ is located 
 well between $T_0$ and $T_0+U$ and additionally 
 $k_{\rm B} T \ll U$, it is 
 \( f_-(T_0) - f_-(T_0+U) \approx 1  \). Then (\ref{eq:numexCF})
 becomes
 \begin{equation}
    \left(\begin{array}{cc} 1 & 1 \\ 1 & 1 \end{array}\right)
    \left(\begin{array}{c} \CF{\n_\up} \\ \CF{\n_\dn} \end{array}\right)
    \approx f_-(T_0) \left(\begin{array}{c} 1 \\ 1 \end{array}\right) ,
 \end{equation}
 with a vanishing determinant of the matrix. Analytically,
 one can strictly show that 
 \( \CF{\n_\up} = \CF{\n_\dn} 
                = f_-(T_0) \left[1+ f_-(T_0)- f_-(T_0+U)\right]^{-1}
 \). The numerical problem can be avoided if the chemical 
 potential is close to an energy pole or if the temperature is high
 enough.

\begin{figure}[ht]    
    \epsfig{file=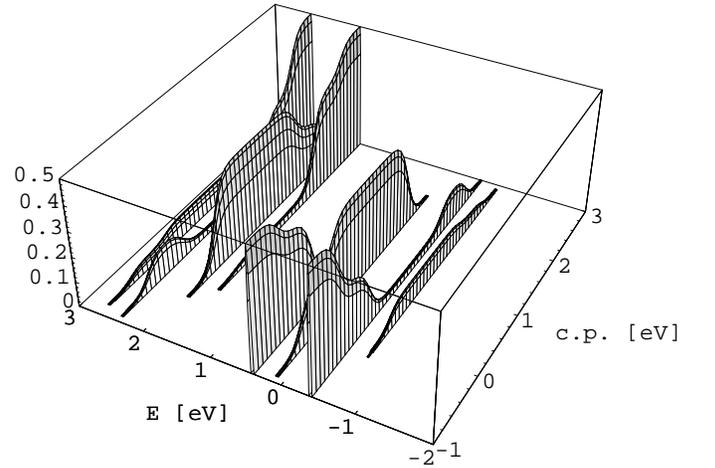, width=0.5\textwidth}
    \caption[Hubbard-cluster - dependence on chemical potential]
    {For the two-site Hubbard-cluster the dependence of the spectral
     weights of $\GF{\cn{i\ps}}{\cd{i\ps}}$ on the chemical 
     potential (c.p.) is given. The energy
     poles are situated at energies $E$ given in appendix 
     \ref{The Hubbard cluster}. The model parameters are: 
     $T_0= J = 0 {\rm\, eV}, \t = 0.4 {\rm\, eV}$ and $U = 2 {\rm\, eV}$.
     A temperature of $T= 100 {\rm\, K}$ is used.
     \label{fg:cp-dep} }
 \end{figure}

 For the two-site Hubbard cluster ${\cal H}_s + {\cal H}_{ss}$ the 
 mathematical systems are much smaller. Hence, already temperatures 
 around $100 {\rm\,K}$ are sufficient to get meaningful results
 (see figure \ref{fg:cp-dep} ) over the whole range of the chemical 
 potential (c.p.). 
 In regions for $\mu$ which are not close to the energy poles all spectral 
 weights remain constant. Figure \ref{fg:cp-dep} also nicely illustrates a further 
 consequence of the particle-hole symmetry. As already mentioned above, the
 energy poles of the one-particle Green's function occur in pairs 
 \( \left( E_k = T_0 + f_k(\t,J,U), \bar E_k = T_0 + U - f_k(-\t,J,U) \right) \).
 We can now see that for the corresponding spectral weights the 
 implication 
 \begin{equation}
   \label{eq:phs-spectr}
   \alpha_k = \alpha_k(\mu,\ps,\t) \,\Longrightarrow\,
   \bar \alpha_k = \alpha_k(U-\mu,-\ps,-\t),
 \end{equation}
 that can be proven analytically, is fulfilled.
 The signs of $\t$ and $\ps$ are not essential because of site and mirror
 symmetry, respectively. Instead, the $\mu$-dependence is the most
 important feature.

\begin{figure}[ht]
    \epsfig{file=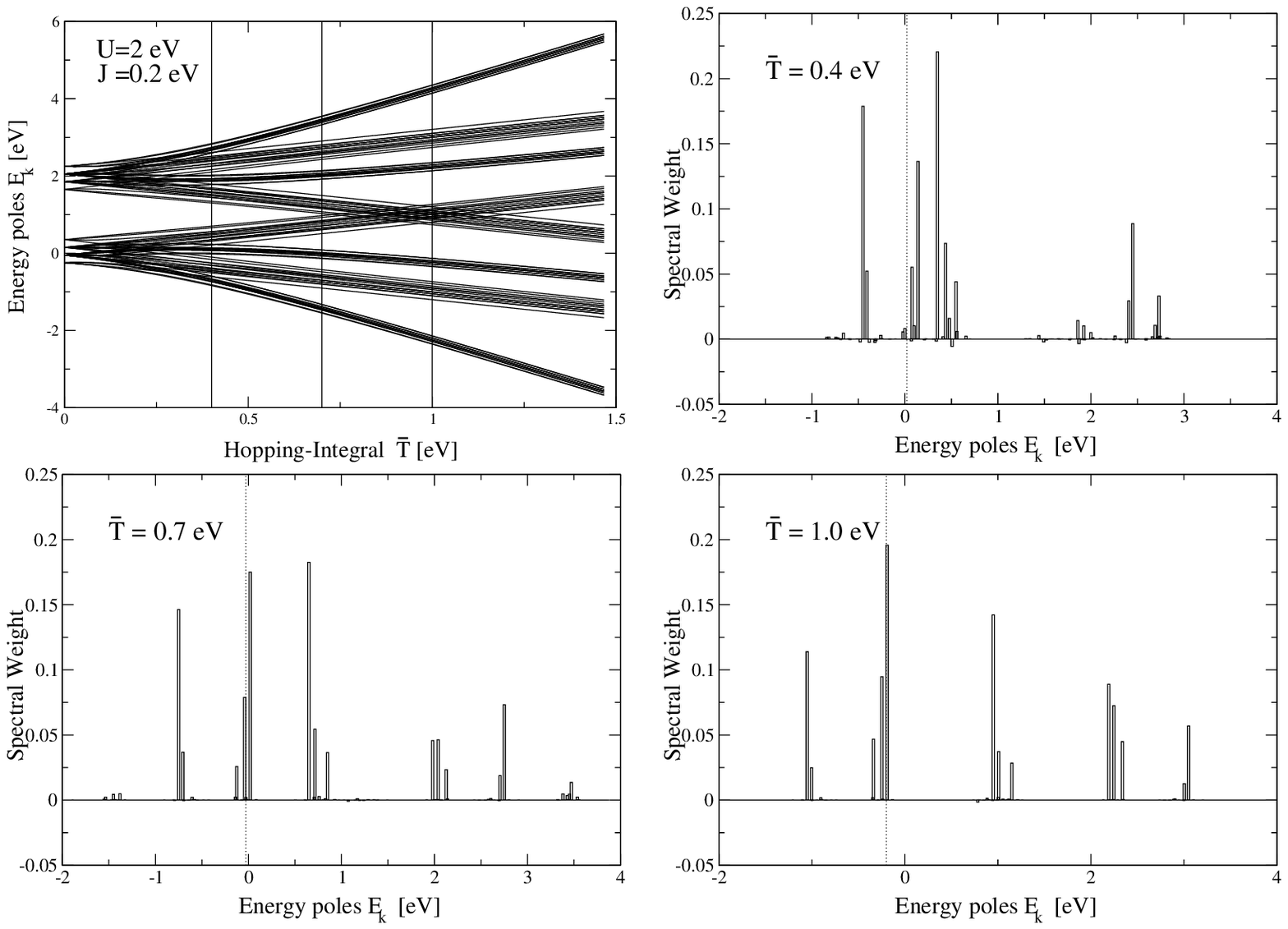, width=0.45\textwidth} 
    \caption[Dependence of spectral weight on hopping integral T]
     {top left: as figure \ref{fg:T-dep2}. others:
     Distribution of the spectral weight of $\GF{\cn{i\ps}}{\cd{i\ps}}$
     for different values of the hopping integral $\t$ (as given). 
     The chemical potential (doted line) is always chosen such, that
     the determinant in (\ref{eq:LGS2}) is not too small. All
     spin-correlation-functions $\CF{\hat S}$ are assumed to be zero.
     The other model parameters are:  $T_0= 0.0 {\rm\, eV}, J = 0.2 {\rm\, eV}, 
      U = 2.0 {\rm\, eV}$. Temperature: $T = 500 {\rm\,K}$. 
     \label{fg:T-dep3}}
 \end{figure}

 Coming back to the complete CKLM we are again especially interested
 in the dependence of the cluster properties on $\t$. For $\t=0$ we
 are able to reproduce the {\it atomic limit} results \cite{NoM84} completely.
 This includes statements on the spectral weights and the fact that
 it vanishes for the four additionally emerged energy poles in
 figure \ref{fg:T-dep1}. The distribution of the spectral weight
 for higher values of $\t$ is shown in figure \ref{fg:T-dep3}. 
 Especially for $\t$ values higher than $0.5 {\rm\, eV}$ the dominance
 of the energy poles of the Hubbard cluster can clearly be seen. 

\begin{figure}[ht]
    \epsfig{file=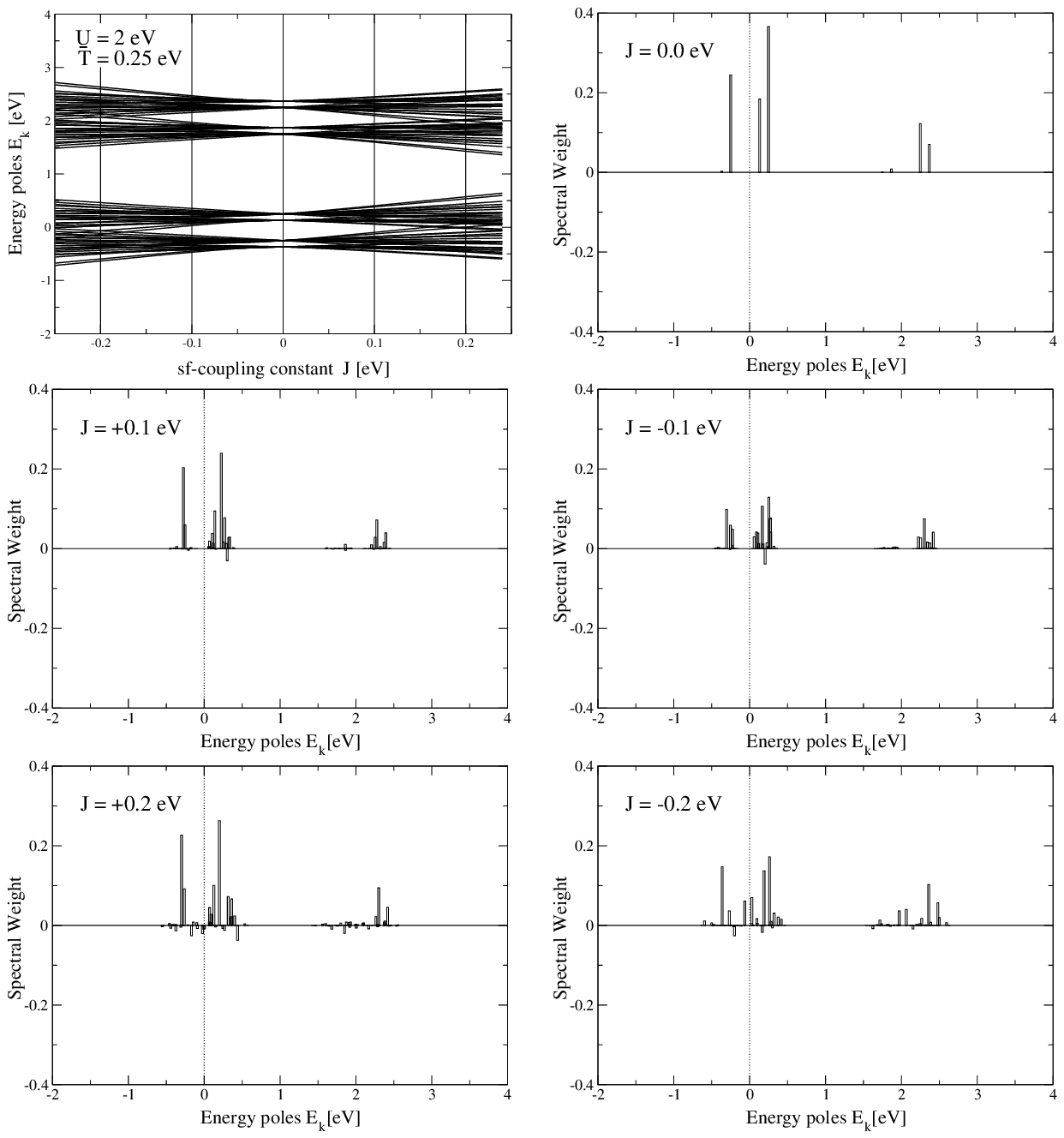, width=0.45\textwidth}  \quad
    \caption[Dependence of spectral weight on coupling constant J]
    {top left: as figure \ref{fg:J-dep1}. others:
     Distribution of the spectral weight of $\GF{\cn{i\up}}{\cd{i\up}}$
     for different values of the sf-coupling constant $J$ (as given)
     with opposite signs. The chemical potential is kept fixed at
     $\mu = 0.0 {\rm\, eV}$. For the spin expectation values the
     specification 
     \( \CF{S^z_1} = \CF{S^z_2} =0.1; \CF{S^z_1 S^z_2} =
             \CF{S^+_1 S^-_2} = \CF{S^-_1 S^+_2} = 0.0 \) 
     has been assumed. The other model parameters are:     
     $T_0= 0.0 {\rm\, eV}, \t = 0.25 {\rm\, eV}, U = 2.0 {\rm\, eV}$. 
     Temperature: $T = 500 {\rm\, K}$. 
     \label{fg:J-dep2}}
 \end{figure}

 The splitting of theses poles and a greater width of the bundles in
 figure \ref{fg:T-dep2} is connected to an increase of $J$.
 Nevertheless, a redistribution of spectral weight happens just 
 gradually, as can be seen in figure \ref{fg:J-dep2}. 
 It is remarkable that up to $|J|=0.25 {\rm\, eV}$ the spectral weight
 is restricted to a few energy poles. One can also see that some of 
 the energy poles get small negative spectral weights. This is caused
 by the two problems mentioned above. On the one hand we have specified
 the values of the spin-correlation functions in figure \ref{fg:J-dep2}.
 The choice does certainly not match with calculations based on
 the two-site Hamiltonian (\ref{eq:Cluster-Ham}). We have stressed 
 above that this might lead to a breakdown of certain symmetries
 of the cluster. The
 fact that every spectral weight must be positive, as can be
 shown with the help of Lehmann's representation, is a property that
 follows from such symmetries. On the other hand we also mentioned
 that an unfavourable choice for the chemical potential $\mu$ 
 implies large numerical errors, due to very small determinants of
 matrices that need to be inverted. This can also cause negative
 values for spectral weights. 
 
\begin{figure}[ht]
    \epsfig{file=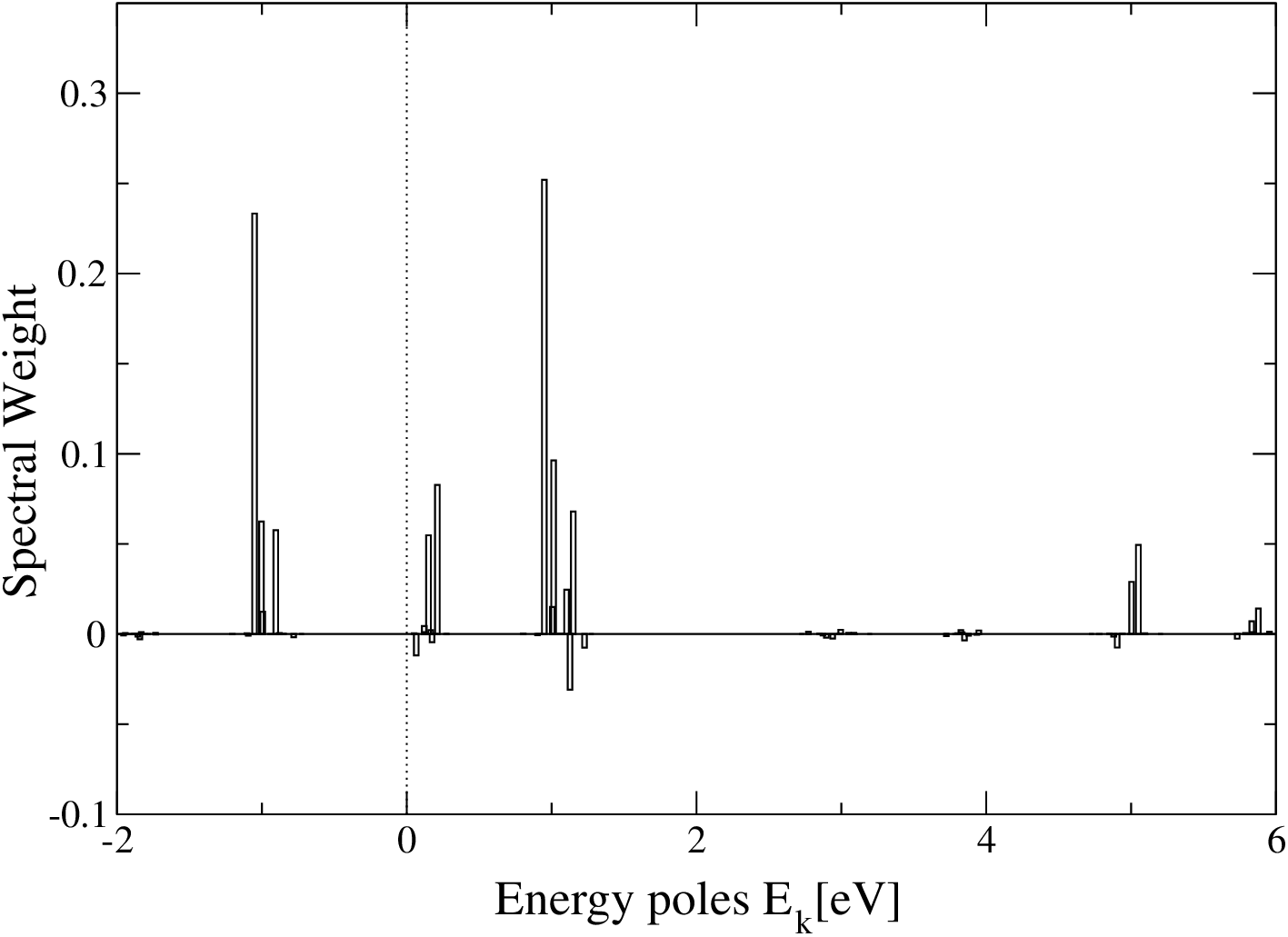, width=0.45\textwidth} 
    \caption[Spectral weights for an antiferromagnetic order]
    {Distribution of the spectral weight over the energy poles
     of \( \GF{\cn{1\up}}{\cd{1\up}} \). By the specification    
     of the spin-correlation functions 
     \( \CF{S^z_1} = 0.1, \CF{S^z_2} = -0.1; \CF{S^z_1 S^z_2} =
             \CF{S^+_1 S^-_2} = \CF{S^-_1 S^+_2} = 0.0 \)
     an antiferromagnetic order is assumed. The model parameters
     are: $T_0= 0.0 {\rm\, eV}, \t = 1.0 {\rm\, eV}, J = 0.2 {\rm\, eV}, 
      U = 4.0 {\rm\, eV}$. Temperature: $T = 500 {\rm\, K}$. 
     \label{fg:afm}}
 \end{figure}

 Our program allows the presentation of the excitation spectrum for 
 various situations. The main intension for the improvement
 of the {\it atomic limit} calculations was the ability to treat
 also an antiferromagnet order of the system of localized spins.
 An example for typical results is given in figure \ref{fg:afm}.
 As in most of the other figures one can see, that even though  
 the Green's function possesses more then 100 energy poles, only
 a few of them have a non-vanishing spectral weight. 
 The expectation of quasi-continuous energy bands, provoked
 by pictures for the energy poles as figure \ref{fg:T-dep1},
 is not met by the result obtained from the calculation of 
 the spectral weights. Like in the {\it atomic limit} we end up
 with a limited number of delta peaks in the spectrum.

\section{Summary}

 We considered the analytical solution for a two-site cluster
 described by the CKLM. In contrast to a foregoing publication
 \cite{HRN01} we did not restrict the occupation of the single
 conduction band. Moreover, a correlation effect in the form of
 a Hubbard interaction in this band is taken into account. 
 The only remaining constraint is $S=1/2$ for the localized spins.

 The mathematical effort to achieve the analytical exact result
 for the Green's function \( \GF{\cn{i\ps}}{\cd{i\ps}} \) is immense.
 It was the goal of the paper to show that it is still manageable and
 to explain the necessary techniques. Major points were the
 separation of the 1040 Green's functions in four density classes,
 the use of intrinsical symmetries and the study of limiting cases
 as the empty conduction band and the Hubbard cluster. 
 As a consequence we found the ``correct'' combinations of 
 Green's functions (appendix \ref{Combinations of Green's functions})
 to obtain system sizes of order $3 \times 3$. 

 The suggested routine for an analytical calculation consists of
 two main parts. First of all the set of equations of motion needs to
 be solved by: Combining their inhomogeneities correctly, 
 including the results for higher density classes as in equation
 (\ref{eq:steps}), inverting the $3 \times 3$ matrices, performing
 partial fraction expansions and undoing the combinations of the
 Green's functions. Secondly, the correlation functions in the 
 spectral weights need to be determined. This happens by: 
 Specifying a desired subset of the correlation functions, 
 applying the spectral theorem to every determined Green's function,
 selecting a representative set of these linear equations, 
 using an LU-decomposition to solve the linear system of equations
 for the unspecified correlation functions. 

 The first part has been done by hand and the analytical expressions
 for the obtained $102$ energy poles as well as a graphical
 presentation of the excitation spectrum has been given in this
 paper. The correlation effect, which leads to a formation of two
 Hubbard bands, and the characteristic dependence on the important
 model parameters can clearly be seen in figure \ref{fg:T-dep1}, 
 \ref{fg:T-dep2} and \ref{fg:J-dep1}. For the second part a computer 
 program has been used, since there exist reliable routines for the 
 numerically exact decomposition of large systems. The resulting vector 
 for the correlation functions was used to discuss the distribution of 
 the spectral weight among the excitation energies. 
 A major conclusion from the provided figures is the concentration
 of the spectral on a few energy poles. The notion of almost 
 ``continuous'' bands formed by $102$ energies is not confirmed
 if the numerators of the Green's functions are considered. 

 We propose to use a subset of the correlation functions in the 
 numerators of the Green's functions to construct a connection of the
 cluster to the lattice. This freedom in the specification of 
 expectation values strongly influences the results for the spectral
 weights. We have pointed out, that as a possible side effect of this
 procedure symmetries of the cluster might be destroyed, leading to
 certain unphysical features. 

 We intend to use the results of this work as a limiting case to test 
 various approximations for the complete CKLM. Furthermore, the cluster 
 result itself can serve as a starting point for an approximation of 
 the lattice. We are convinced that techniques similar to what has 
 been proposed here can be used to treat slightly more complex systems,
 like a cluster with a higher value for the localized spin $S$ or another
 number of lattice sites. 
 
\section*{Acknowledgments}
 One of us (T.~H.) gratefully acknowledges the financial support of the 
 {\it Studienstiftung des deutschen Volkes}. This work also benefitted from the
 financial support of the {\it Sonderforschungsbereich SFB 290} of the 
 Deutsche Forschungsgemeinschaft.

\newpage
\begin{widetext}
\begin{appendix}
 \section{The Hubbard cluster}
 \label{The Hubbard cluster}
 
  An analytical solution of a two-site Hubbard cluster has already 
  been given by several authors \cite{Sch01,AMS01,Mat99}. We present
  here a direct solution, that is compatible to the language
  of equations of motion used in the rest of the paper. 

  We study the Hamiltonian
  \begin{equation}
     \label{eq:Hub_Ham}
     {\cal H} = \sum_\ps \left\{
                \t \left( \cd{1\ps} \cn{2\ps} + \cd{2\ps} \cn{1\ps} \right)
              + \frac{U}{2} \sum_{k=1}^2 \n_{k\ps} \n_{k\ms} \right\}.
  \end{equation}
  Using abbreviation (\ref{eq:comb1}) we need to consider a
  set of Green's functions which is almost identical to the list
  given in section \ref{Solution for the equations of motion}. 
  However, in contrast to the CKLM cluster we have \( \hat S \equiv 1 \)
  what decreases the number of functions drastically.
  We are left with the following set of equations of motion, 
  written again in a matrix representation:
  {\setlength{\arraycolsep}{0pt}
  \begin{equation} 
  E \left( \begin{array}{c}
      G_{\rm A} \\ G_{\rm B} \\ G_{\rm C} \\ G_{\rm D} \\ G_{\rm F} \\ 
      G_{\rm G} \\ G_{\rm H} \\ G_{\rm J} \\ G_{\rm K} \\ G_{\rm L} \\ G_{\rm M} 
    \end{array} \right)
  - \left( \begin{array}{c|cccc|c|cccc|c}
     \vz\t & U &&&&&&&& \\ \hline
        & U & \vz\t & \vz\t & -\vz\t &&&&&& \\
        & \vz\t &    & -\vz\t & \vz\t & U &&&&& \\
        & \vz\t & -\vz\t & & \vz\t & & U &&&& \\
        & -\vz\t & \vz\t & \vz\t & U & & & -U &&& \\ \hline
        &&&&& \vz\t+U &&&&& \\ \hline
        &&&&&& U & \vz\t & \vz\t & -\vz\t & \\
        &&&&&& \vz\t && -\vz\t & \vz\t & \\
        &&&&&& \vz\t & -\vz\t & U & \vz\t &\\
        &&&&&& -\vz\t & \vz\t & \vz\t && U \\ \hline
        &&&&&&&&&& \vz\t+U 
    \end{array} \right)
    \left( \begin{array}{c}
      G_{\rm A} \\ G_{\rm B} \\ G_{\rm C} \\ G_{\rm D} \\ G_{\rm F} \\ 
      G_{\rm G} \\ G_{\rm H} \\ G_{\rm J} \\ G_{\rm K} \\ G_{\rm L} \\ G_{\rm M} 
    \end{array} \right)
  = \left( \begin{array}{c}
      K_{\rm A} \\ K_{\rm B} \\ K_{\rm C} \\ K_{\rm D} \\ K_{\rm F} \\ 
      K_{\rm G} \\ K_{\rm H} \\ K_{\rm J} \\ K_{\rm K} \\ K_{\rm L} \\ K_{\rm M} 
    \end{array} \right) \\
  \end{equation}}

  It is now essential to introduce the combinations
   \begin{equation}
     \label{eq:Hub_Komb}
     \begin{array}{rclrcl}
        X_{\rm b}^{(\vz)} &=& X_{\rm B}^{(\vz)} + X_{\rm F}^{(\vz)} ,\hspace{2em} &
        X_{\rm f}^{(\vz)} &=& X_{\rm B}^{(\vz)} - X_{\rm F}^{(\vz)} ,\\
        X_{\rm c}^{(\vz)} &=& X_{\rm C}^{(\vz)} + X_{\rm D}^{(\vz)} ,&
        X_{\rm d}^{(\vz)} &=& X_{\rm C}^{(\vz)} - X_{\rm D}^{(\vz)} ,\\[1ex]
        X_{\rm k}^{(\vz)} &=& X_{\rm K}^{(\vz)} + X_{\rm H}^{(\vz)} ,&
        X_{\rm h}^{(\vz)} &=& X_{\rm K}^{(\vz)} - X_{\rm H}^{(\vz)} ,\\
        X_{\rm l}^{(\vz)} &=& X_{\rm L}^{(\vz)} + X_{\rm J}^{(\vz)} ,&
        X_{\rm j}^{(\vz)} &=& X_{\rm L}^{(\vz)} - X_{\rm J}^{(\vz)} ,
     \end{array}
   \end{equation}
  where ``$X$'' stands for ``$G$'' or ``$K$''.
  As one can see the matrix representation of the original equations 
  of motion contains blocks of the size $4 \times 4$. After the proposed combination
  the remaining maximum block size is $2 \times 2$, which can easily be inverted:
  \begin{equation}
    \label{eq:Hub_22inverse}
     \left( \begin{array}{cc}
               \EM-U & -2\t \\
               - 2\t & \EM 
            \end{array} \right)^{-1}  \!\!
     = \sum_{\gamma= \pm 1} 
        \frac{ \frac{1}{2} \mbox{\small \( 
               \left(\begin{array}{cc} 1 & 0 \\ 0 & 1 \end{array}\right) \)}
             + \gamma \frac{1}{4 b} \mbox{\small \( 
               \left( \begin{array}{cc} U & 4\t \\ 4\t & - U
               \end{array} \right) \)}
             }{\EM-\Uh-\gamma  b}  
  \end{equation}
  The form of the energy constant
  \(   b = \sqrt{ \left(\Uh\right)^2 + 4 \t^2 } \)  
  demonstrates, that quadratic system sizes are inevitable.  
  We get the following matrix for the equations of motion:
  {\setlength{\arraycolsep}{0pt}
  \begin{equation} 
   \label{eq:Hub_matrix2}
  E \left( \begin{array}{c}
      G_{\rm A}\\ G_{\rm b} \\ G_{\rm c} \\ G_{\rm f} \\ G_{\rm d} \\ G_{\rm G}\\
      G_{\rm h} \\ G_{\rm j} \\ G_{\rm k} \\ G_{\rm l} \\ G_{\rm M}
    \end{array} \right)
  - \left( \begin{array}{c|cccc|c|cccc|c}
     \vz\t & \Uh && \Uh &&&&&&& \\ \hline
        & U\!-\!\vz\t & 2\vz\t &&&&& \Uh && -\Uh & \\ 
        & 2\vz\t & -\vz\t &&& U & -\Uh && \Uh && \\
        &&& U\!+\!\vz\t &&&& -\Uh && \Uh &\\
        &&&& \vz\t & U & \Uh && -\Uh &&\\ \hline
        &&&&& \vz\t+U &&&&& \\ \hline
        &&&&&& U\!-\!\vz\t & 2\vz\t &&& \\
        &&&&&& 2\vz\t & -\vz\t &&& U \\ 
        &&&&&&&& U\!+\!\vz\t &&  \\
        &&&&&&&&& \vz\t & U \\ \hline
        &&&&&&&&&& \vz\t+U
    \end{array} \right)
    \left( \begin{array}{c}
      G_{\rm A}\\ G_{\rm b} \\ G_{\rm c} \\ G_{\rm f} \\ G_{\rm d} \\ G_{\rm G}\\
      G_{\rm h} \\ G_{\rm j} \\ G_{\rm k} \\ G_{\rm l} \\ G_{\rm M}
    \end{array} \right)
  = \left( \begin{array}{c}
      K_{\rm A}\\ K_{\rm b} \\ K_{\rm c} \\ K_{\rm f} \\ K_{\rm d} \\ K_{\rm G}\\
      K_{\rm h} \\ K_{\rm j} \\ K_{\rm k} \\ K_{\rm l} \\ K_{\rm M}
    \end{array} \right) 
  \end{equation}}

  What remains is a straightforward solution of (\ref{eq:Hub_matrix2}).
  The expression of $G_{\rm M}^{(\vz)}$ can immediately be written down.
  To get $G_{\rm l}^{(\vz)}$ and $G_{\rm k}^{(\vz)}$ in a form with linear energy poles,
  a single partial fraction expansion is necessary. For $G_{\rm h}^{(\vz)}$
  and $G_{\rm j}^{(\vz)}$ one has to apply (\ref{eq:Hub_22inverse}). And with
  a few more steps one gets\\
  \begin{eqnarray}
    G_{\rm A}^{(\vz)}
    &=& \frac{K_{\rm A} }{E-\vz\t}
     -  \frac{K_{\rm b}^{(\vz)}+K_{\rm c}^{(\vz)}+K_{\rm f}^{(\vz)}}
             {2(E-\vz\t)}
     +  \frac{K_{\rm S}^{(\vz)}}{2(E-\vz\t)}
     +  \frac{K_{\rm f}^{(\vz)} + K_{\rm S}^{(\vz)}}{2(E-\vz\t-U)} \\
    &+& \sum_{\gamma=\pm 1} \left\{
         \frac{\left(1 + \gamma\frac{2\vz\t}{b} + \gamma\frac{U}{2b}\right) 
               K_{\rm b}^{(\vz)} 
       + \left(1 + \gamma\frac{2\vz\t}{b} - \gamma\frac{U}{2b}\right) 
               K_{\rm c}^{(\vz)} }{ 4 (E+\vz\t-\Uh-\gamma b) }
       - \frac{ \left(1+\gamma\frac{2\vz\t}{b} \right) K_{\rm S}^{(\vz)} }
              { 2 (E+\vz\t-\Uh-\gamma b) } 
        \right\} . \nonumber \\ \nonumber
  \end{eqnarray}
  We have defined the sum of correlation functions
  \begin{equation}
    \label{eq:Hub_sumCF}
     K_{\rm S}^{(\vz)}
        = \vz\CF{\n_{\os\ps} \cd{\os\ms} \cn{\ts\ms}}
        -    \CF{\cd{\ts\ms} \cn{\os\ms} \cd{\ts\ps} \cn{\os\ps}}
        + \vz\CF{\n_{\os\ps} \cd{\ts\ms} \cn{\os\ms}} 
        -    \CF{\cd{\os\ms} \cn{\ts\ms} \cd{\ts\ps} \cn{\os\ps}} 
        +    \CF{\n_{\os\ms} \n_{\ts\ms}} .
  \end{equation}
  As a last step on the way to the desired single-particle Green's 
  function one has to undo the combination (\ref{eq:comb1}) by
  considering the sum 
  \( \frac{1}{2} \left( G_{\rm A}^{(+)} + G_{\rm A}^{(-)} \right) \).
  Then we obtain the result\\
  \begin{eqnarray*}
     \GF{\cn{\ts\ps}}{\cd{\ts\ps}} 
       &=& \frac{2 - K_{\rm b}^{(+)}- K_{\rm c}^{(+)}- K_{\rm f}^{(+)}
                +  K_{\rm S}^{(+)} }{4(E-\t)}
        +  \frac{K_{\rm f}^{(+)} + K_{\rm S}^{(+)}}{4(E-\t-U)} \\
       &+& \frac{2 - K_{\rm b}^{(-)}- K_{\rm c}^{(-)}- K_{\rm f}^{(-)}
                +  K_{\rm S}^{(-)} }{4(E+\t)}
        +  \frac{K_{\rm f}^{(-)} + K_{\rm S}^{(-)}}{4(E+\t-U)} \\
       &+& \frac{ \left(1 + \frac{2\t}{ b} \right)
                  (K_{\rm b}^{(+)} + K_{\rm c}^{(+)} - 2K_{\rm S}^{(+)})
                + \frac{U}{2 b} (K_{\rm b}^{(+)} - K_{\rm c}^{(+)})
                }{ 8 (E+\t-\Uh- b) } \\
       &+& \frac{ \left(1 -\frac{2\t}{ b} \right) 
                  (K_{\rm b}^{(+)} + K_{\rm c}^{(+)} - 2K_{\rm S}^{(+)})
                - \frac{U}{2 b} (K_{\rm b}^{(+)} - K_{\rm c}^{(+)})
                }{ 8 (E+\t-\Uh+ b) } \\
       &+& \frac{ \left(1 - \frac{2\t}{ b} \right) 
                  (K_{\rm b}^{(-)} + K_{\rm c}^{(-)} - 2K_{\rm S}^{(-)})
                + \frac{U}{2 b} (K_{\rm b}^{(-)} - K_{\rm c}^{(-)})
                }{ 8 (E-\t-\Uh- b) } \\
       &+& \frac{ \left(1 + \frac{2\t}{ b} \right) 
                  (K_{\rm b}^{(-)} + K_{\rm c}^{(-)} - 2K_{\rm S}^{(-)})
                - \frac{U}{2 b} (K_{\rm b}^{(-)} - K_{\rm c}^{(-)})
                }{ 8 (E-\t-\Uh+ b) } .
  \end{eqnarray*}

 \section{Combinations of Green's functions}
 \label{Combinations of Green's functions}

  First of all the notation for the Green's functions needs to
  be clarified. Each of them gets to subindex that consists of
  a number and a letter. The latter part has already been introduced 
  above when the list of Green's functions with different 
  possibilities for the fermionic operators is given. 
  The former part enumerates the possibilities for spin operators
  $\hat S$. The numbers $ 1, \ldots, 10$ are used for $ \hat S =
    1, S_\ts^z, S_\os^z, S_\ts^{-\ps}, S_\os^{-\ps}, S_\ts^z S_\os^z,
    S_\os^z S_\ts^{-\ps}, S_\ts^z S_\os^{-\ps}, S_\os^{-\ps} S_\ts^{\ps},
    S_\ts^{-\ps} S_\os^{\ps}$.
  Additionally, the Green's functions 
  \( \Gf{\hat S \cd{\os\ms} \cn{\os\ps} \cn{\ts\ps}}_{Nx}, \) and
  \( \Gf{\hat S \n_{\ts\ps} \cd{\os\ps} \cn{\os\ms} \cn{\ts\ms}}_{Px} \)
  with $x=1,\ldots,5$ for 
  \(  \hat S = S^\ps_\ts, S^\ps_\os, S^z_\os S^\ps_\ts, S^z_\ts S^\ps_\os,
               S^\ms_\os S^\ms_\ts \)
  appear in the necessary equations of motion. Each Green's function
  taken $8$ times for the different possibilities for $\ts, \ps$ and
  $\vz$ gives the number $1040$. 
 

  The first step of combinations is determined by the spin operators.
  We use the rules that have been applied to the insulator \cite{HRN01}.
  \begin{eqnarray*}
    H_{2\ldots}^{(\pm)} &=& 
            \mbox{$\frac{\hbar^2}{4}$} G_{1\ldots} \pm \mbox{$\frac{\hbar}{2}$} G_{2\ldots}
        \pm \mbox{$\frac{\hbar}{2}$} G_{3\ldots} +       G_{6\ldots} \\
    H_{3\ldots}^{(\pm)} &=& 
            \mbox{$\frac{\hbar^2}{4}$} G_{1\ldots} \mp \mbox{$\frac{\hbar}{2}$} G_{2\ldots}
        \pm \mbox{$\frac{\hbar}{2}$} G_{3\ldots} -       G_{6\ldots} \\
    H_{4\ldots}^{(\pm)} &=& 
            \mbox{$\frac{\hbar}{2}$} G_{4\ldots} \pm     G_{7\ldots} \\
    H_{5\ldots}^{(\pm)} &=& 
            \mbox{$\frac{\hbar}{2}$} G_{5\ldots} \pm     G_{8\ldots} \\
    H_{N1}^{(\pm)} &=& \mbox{$\frac{\hbar}{2}$} G_{N1} \pm G_{N3} \qquad
    H_{P1}^{(\pm)} \,=\,  \mbox{$\frac{\hbar}{2}$} G_{P1} \pm G_{P3} \\
    H_{N2}^{(\pm)} &=& \mbox{$\frac{\hbar}{2}$} G_{N2} \pm G_{N4} \qquad
    H_{P2}^{(\pm)} \,=\,  \mbox{$\frac{\hbar}{2}$} G_{P2} \pm G_{P4} 
  \end{eqnarray*}

  The next step produces the correct combinations of fermionic operators
  of different Green's functions. We use the rules (\ref{eq:Hub_Komb})
  applied to the Hubbard cluster. 

  \subsection*{Density class $\rmII$}
  \hfill
  \begin{minipage}{5cm}
   \begin{eqnarray*}
     H_{4l}^{(-)} &=& H_{4L}^{(-)} + H_{5J}^{(-)} \\
     H_{4k}^{(-)} &=& H_{4K}^{(-)} + H_{5H}^{(-)} \\
     H_{5h}^{(-)} &=& H_{4K}^{(-)} - H_{5H}^{(-)} \\
     H_{5j}^{(-)} &=& H_{4L}^{(-)} - H_{5J}^{(-)} 
   \end{eqnarray*}
  \end{minipage}
  \begin{minipage}{5cm}
   \begin{eqnarray*}
     H_{5l}^{(-)} &=& H_{5L}^{(-)} + H_{4J}^{(-)} \\
     H_{5k}^{(-)} &=& H_{5K}^{(-)} + H_{4H}^{(-)} \\
     H_{4h}^{(-)} &=& H_{5K}^{(-)} - H_{4H}^{(-)} \\
     H_{4j}^{(-)} &=& H_{5L}^{(-)} - H_{4J}^{(-)} 
   \end{eqnarray*}
  \end{minipage}
  \begin{minipage}{5cm}
   \begin{eqnarray*}
     H_{2l}^{(\pm)} &=& H_{2L}^{(\pm)} + H_{2J}^{(\pm)} \\
     H_{2k}^{(\pm)} &=& H_{2K}^{(\pm)} + H_{2H}^{(\pm)} \\
     H_{2h}^{(\pm)} &=& H_{2K}^{(\pm)} - H_{2H}^{(\pm)} \\
     H_{2j}^{(\pm)} &=& H_{2L}^{(\pm)} - H_{2J}^{(\pm)} 
   \end{eqnarray*}
  \end{minipage}
  \hfill { }

  \hfill
  \begin{minipage}{5cm}
   \begin{eqnarray*}
     H_{3g}^{(-)} &=& H_{3G}^{(-)} + G_{10G} \\
     H_{3g}^{(+)} &=& H_{3G}^{(+)} + G_{ 9G} \\[1ex]
     H_{6g}^{(-)} &=& H_{3G}^{(-)} - G_{10G} \\
     H_{6g}^{(+)} &=& H_{3G}^{(+)} - G_{ 9G} 
   \end{eqnarray*}
  \end{minipage}
  \begin{minipage}{5cm}
   \begin{eqnarray*}
     H_{4g}^{(+)} &=& H_{4G}^{(+)} + H_{5G}^{(+)} \\
     H_{5g}^{(+)} &=& H_{4G}^{(+)} - H_{5G}^{(+)} \\
     H_{4g}^{(*)} &=& H_{4G}^{(-)} + H_{P2}^{(+)} \\
     H_{5g}^{(*)} &=& H_{5G}^{(-)} + H_{P1}^{(+)} \\
     H_{2g}^{(+)} &=& H_{2G}^{(+)} + G_{P5}    
   \end{eqnarray*}
  \end{minipage}
  \begin{minipage}{5cm}
   \begin{eqnarray*}
     H_{p1}^{(-)} &=& H_{P1}^{(-)} + H_{P2}^{(-)} \\
     H_{p2}^{(-)} &=& H_{P1}^{(-)} - H_{P2}^{(-)} \\
     H_{p2}^{(*)} &=& H_{4G}^{(-)} - H_{P2}^{(+)} \\
     H_{p1}^{(*)} &=& H_{5G}^{(-)} - H_{P1}^{(+)} \\
     H_{p5}^{(+)} &=& H_{2G}^{(+)} - G_{P5} 
   \end{eqnarray*}
  \end{minipage}
  \hfill { }\\
  \hfill
  \begin{minipage}{0.48\textwidth}
   \begin{eqnarray*}
     H_{4j+5j}^{(++)} &=& \big( H_{4L}^{(+)} - H_{4J}^{(+)} \big) 
                        + \big( H_{5L}^{(+)} - H_{5J}^{(+)} \big) \\
     H_{4h+5h}^{(++)} &=& \big( H_{4K}^{(+)} - H_{4H}^{(+)} \big) 
                        + \big( H_{5K}^{(+)} - H_{5H}^{(+)} \big) \\
     H_{4h-5h}^{(++)} &=& \big( H_{4K}^{(+)} - H_{4H}^{(+)} \big) 
                        - \big( H_{5K}^{(+)} - H_{5H}^{(+)} \big) \\
     H_{6l-6j}^{(-+)} &=& \big( H_{3J}^{(-)} - G_{ 9L} \big) 
                        - \big( H_{3L}^{(+)} - G_{10J} \big) \\
     H_{6h+6k}^{(-+)} &=& \big( H_{3K}^{(-)} - G_{ 9H} \big) 
                        + \big( H_{3H}^{(+)} - G_{10K} \big) \\
     H_{6h-6k}^{(-+)} &=& \big( H_{3K}^{(-)} - G_{ 9H} \big) 
                        - \big( H_{3H}^{(+)} - G_{10K} \big) \\
     H_{6k+6h}^{(-+)} &=& \big( H_{3H}^{(-)} - G_{ 9K} \big) 
                        + \big( H_{3K}^{(+)} - G_{10H} \big) \\
     H_{6k-6h}^{(-+)} &=& \big( H_{3H}^{(-)} - G_{ 9K} \big) 
                        - \big( H_{3K}^{(+)} - G_{10H} \big) \\
     H_{6j-6l}^{(-+)} &=& \big( H_{3L}^{(-)} - G_{ 9J} \big) 
                        - \big( H_{3J}^{(+)} - G_{10L} \big) \\
     H_{4j-5j}^{(++)} &=& \big( H_{4L}^{(+)} - H_{4J}^{(+)} \big) 
                        - \big( H_{5L}^{(+)} - H_{5J}^{(+)} \big) \\
     H_{6l+6j}^{(-+)} &=& \big( H_{3J}^{(-)} - G_{ 9L} \big) 
                        + \big( H_{3L}^{(+)} - G_{10J} \big) \\
     H_{6j+6l}^{(-+)} &=& \big( H_{3L}^{(-)} - G_{ 9J} \big) 
                        + \big( H_{3J}^{(+)} - G_{10L} \big) 
   \end{eqnarray*}
  \end{minipage}
  \begin{minipage}{0.48\textwidth}
   \begin{eqnarray*}
     H_{4k+5k}^{(++)} &=& \big( H_{4K}^{(+)} + H_{4H}^{(+)} \big) 
                        + \big( H_{5K}^{(+)} + H_{5H}^{(+)} \big) \\
     H_{4k-5k}^{(++)} &=& \big( H_{4K}^{(+)} + H_{4H}^{(+)} \big) 
                        - \big( H_{5K}^{(+)} + H_{5H}^{(+)} \big) \\
     H_{4l-5l}^{(++)} &=& \big( H_{4L}^{(+)} + H_{4J}^{(+)} \big) 
                        - \big( H_{5L}^{(+)} + H_{5J}^{(+)} \big) \\
     H_{3j-3l}^{(-+)} &=& \big( H_{3J}^{(-)} + G_{ 9L} \big) 
                        - \big( H_{3L}^{(+)} + G_{10J} \big) \\
     H_{3h+3k}^{(-+)} &=& \big( H_{3H}^{(-)} + G_{ 9K} \big) 
                        + \big( H_{3K}^{(+)} + G_{10H} \big) \\
     H_{3h-3k}^{(-+)} &=& \big( H_{3H}^{(-)} + G_{ 9K} \big) 
                        - \big( H_{3K}^{(+)} + G_{10H} \big) \\
     H_{3k+3h}^{(-+)} &=& \big( H_{3K}^{(-)} + G_{ 9H} \big) 
                        + \big( H_{3H}^{(+)} + G_{10K} \big) \\
     H_{3k-3h}^{(-+)} &=& \big( H_{3K}^{(-)} + G_{ 9H} \big) 
                        - \big( H_{3H}^{(+)} + G_{10K} \big) \\
     H_{3l-3j}^{(-+)} &=& \big( H_{3L}^{(-)} + G_{ 9J} \big) 
                        - \big( H_{3J}^{(+)} + G_{10L} \big) \\
     H_{4l+5l}^{(++)} &=& \big( H_{4L}^{(+)} + H_{4J}^{(+)} \big) 
                        + \big( H_{5L}^{(+)} + H_{5J}^{(+)} \big) \\
     H_{3j+3l}^{(-+)} &=& \big( H_{3J}^{(-)} + G_{ 9L} \big) 
                        + \big( H_{3L}^{(+)} + G_{10J} \big) \\
     H_{3l+3j}^{(-+)} &=& \big( H_{3L}^{(-)} + G_{ 9J} \big) 
                        + \big( H_{3J}^{(+)} + G_{10L} \big) 
   \end{eqnarray*}
  \end{minipage}
  \hfill { }\\[-3ex]

  \subsection*{Density class $\rmIII$}
  \vspace{-2ex}
  \hfill
  \begin{minipage}{5cm}
   \begin{eqnarray*}
     H_{4b}^{(+)} &=& H_{4B}^{(+)} + H_{5F}^{(+)} \\
     H_{4c}^{(+)} &=& H_{4C}^{(+)} + H_{5D}^{(+)} \\
     H_{5d}^{(+)} &=& H_{4C}^{(+)} - H_{5D}^{(+)} \\
     H_{5f}^{(+)} &=& H_{4B}^{(+)} - H_{5F}^{(+)} 
   \end{eqnarray*}
  \end{minipage}
  \begin{minipage}{5cm}
   \begin{eqnarray*}
   H_{5b}^{(+)} &=& H_{5B}^{(+)} + H_{4F}^{(+)} \\
   H_{5c}^{(+)} &=& H_{5C}^{(+)} + H_{4D}^{(+)} \\
   H_{4d}^{(+)} &=& H_{5C}^{(+)} - H_{4D}^{(+)} \\
   H_{4f}^{(+)} &=& H_{5B}^{(+)} - H_{4F}^{(+)} 
 \end{eqnarray*}
  \end{minipage}
  \begin{minipage}{5cm}
   \begin{eqnarray*}
     H_{2b}^{(\pm)} &=& H_{2B}^{(\pm)} + H_{2F}^{(\pm)} \\
     H_{2c}^{(\pm)} &=& H_{2C}^{(\pm)} + H_{2D}^{(\pm)} \\
     H_{2d}^{(\pm)} &=& H_{2C}^{(\pm)} - H_{2D}^{(\pm)} \\
     H_{2f}^{(\pm)} &=& H_{2B}^{(\pm)} - H_{2F}^{(\pm)} 
   \end{eqnarray*}
  \end{minipage}
  \hfill { } \\[-2ex]

  \hfill
  \begin{minipage}{5cm}
   \begin{eqnarray*}
     H_{3r}^{(+)} &=& H_{3R}^{(+)} + G_{10R} \\
     H_{3r}^{(-)} &=& H_{3R}^{(-)} + G_{ 9R} \\[1ex]
     H_{6r}^{(+)} &=& H_{3R}^{(+)} - G_{10R} \\
     H_{6r}^{(-)} &=& H_{3R}^{(-)} - G_{ 9R} 
   \end{eqnarray*}
  \end{minipage}
  \begin{minipage}{5cm}
   \begin{eqnarray*}
     H_{4r}^{(-)} &=& H_{4R}^{(-)} + H_{5R}^{(-)} \\
     H_{5r}^{(-)} &=& H_{4R}^{(-)} - H_{5R}^{(-)} \\
     H_{4r}^{(*)} &=& H_{4R}^{(+)} + H_{N2}^{(-)} \\
     H_{5r}^{(*)} &=& H_{5R}^{(+)} + H_{N1}^{(-)} \\
     H_{2r}^{(-)} &=& H_{2R}^{(-)} + G_{N5}    
   \end{eqnarray*}
  \end{minipage}
  \begin{minipage}{5cm}
   \begin{eqnarray*}
     H_{n1}^{(+)} &=& H_{N1}^{(+)} + H_{N2}^{(+)} \\
     H_{n2}^{(+)} &=& H_{N1}^{(+)} - H_{N2}^{(+)} \\
     H_{n2}^{(*)} &=& H_{4R}^{(+)} - H_{N2}^{(-)} \\
     H_{n1}^{(*)} &=& H_{5R}^{(+)} - H_{N1}^{(-)} \\
     H_{n5}^{(-)} &=& H_{2R}^{(-)} - G_{N5} 
   \end{eqnarray*}
  \end{minipage}
  \hfill { }\\[-1ex]
  \hfill
  \begin{minipage}{0.48\textwidth}
   \begin{eqnarray*}
     H_{4b+5b}^{(--)} &=& \big( H_{4B}^{(-)} + H_{4F}^{(-)} \big)
                        + \big( H_{5B}^{(-)} + H_{5F}^{(-)} \big) \\
     H_{4c+5c}^{(--)} &=& \big( H_{4C}^{(-)} + H_{4D}^{(-)} \big)
                        + \big( H_{5C}^{(-)} + H_{5D}^{(-)} \big) \\
     H_{4c-5c}^{(--)} &=& \big( H_{4C}^{(-)} + H_{4D}^{(-)} \big)
                        - \big( H_{5C}^{(-)} + H_{5D}^{(-)} \big) \\
     H_{3b+3f}^{(+-)} &=& \big( H_{3B}^{(+)} + G_{ 9F} \big)
                        + \big( H_{3F}^{(-)} + G_{10B} \big) \\
     H_{3c+3d}^{(+-)} &=& \big( H_{3C}^{(+)} + G_{ 9D} \big)
                        + \big( H_{3D}^{(-)} + G_{10C} \big) \\
     H_{3c-3d}^{(+-)} &=& \big( H_{3C}^{(+)} + G_{ 9D} \big)
                        - \big( H_{3D}^{(-)} + G_{10C} \big) \\
     H_{3c+3d}^{(-+)} &=& \big( H_{3C}^{(-)} + G_{10D} \big)
                        + \big( H_{3D}^{(+)} + G_{ 9C} \big) \\
     H_{3c-3d}^{(-+)} &=& \big( H_{3C}^{(-)} + G_{10D} \big)
                        - \big( H_{3D}^{(+)} + G_{ 9C} \big) \\
     H_{3b+3f}^{(-+)} &=& \big( H_{3B}^{(-)} + G_{10F} \big)
                        + \big( H_{3F}^{(+)} + G_{ 9B} \big) \\
     H_{4b-5b}^{(--)} &=& \big( H_{4B}^{(-)} + H_{4F}^{(-)} \big)
                        - \big( H_{5B}^{(-)} + H_{5F}^{(-)} \big) \\
     H_{3b-3f}^{(-+)} &=& \big( H_{3B}^{(-)} + G_{10F} \big)
                        - \big( H_{3F}^{(+)} + G_{ 9B} \big) \\
     H_{3b-3f}^{(+-)} &=& \big( H_{3B}^{(+)} + G_{ 9F} \big)
                        - \big( H_{3F}^{(-)} + G_{10B} \big) \\
   \end{eqnarray*}
  \end{minipage}
  \begin{minipage}{0.48\textwidth}
   \begin{eqnarray*}
     H_{4d+5d}^{(--)} &=& \big( H_{4C}^{(-)} - H_{4D}^{(-)} \big)
                        + \big( H_{5C}^{(-)} - H_{5D}^{(-)} \big) \\
     H_{4d-5d}^{(--)} &=& \big( H_{4C}^{(-)} - H_{4D}^{(-)} \big)
                        - \big( H_{5C}^{(-)} - H_{5D}^{(-)} \big) \\
     H_{4f-5f}^{(--)} &=& \big( H_{4B}^{(-)} - H_{4F}^{(-)} \big)
                        - \big( H_{5B}^{(-)} - H_{5F}^{(-)} \big) \\
     H_{6b+6f}^{(+-)} &=& \big( H_{3F}^{(+)} - G_{ 9B} \big)
                        + \big( H_{3B}^{(-)} - G_{10F} \big) \\
     H_{6c+6d}^{(+-)} &=& \big( H_{3D}^{(+)} - G_{ 9C} \big)
                        + \big( H_{3C}^{(-)} - G_{10D} \big) \\
     H_{6c-6d}^{(+-)} &=& \big( H_{3D}^{(+)} - G_{ 9C} \big)
                        - \big( H_{3C}^{(-)} - G_{10D} \big) \\
     H_{6c+6d}^{(-+)} &=& \big( H_{3D}^{(-)} - G_{10C} \big)
                        + \big( H_{3C}^{(+)} - G_{ 9D} \big) \\
     H_{6c-6d}^{(-+)} &=& \big( H_{3D}^{(-)} - G_{10C} \big)
                        - \big( H_{3C}^{(+)} - G_{ 9D} \big) \\
     H_{6b+6f}^{(-+)} &=& \big( H_{3F}^{(-)} - G_{10B} \big)
                        + \big( H_{3B}^{(+)} - G_{ 9F} \big) \\
     H_{4f+5f}^{(--)} &=& \big( H_{4B}^{(-)} - H_{4F}^{(-)} \big)
                        + \big( H_{5B}^{(-)} - H_{5F}^{(-)} \big) \\
     H_{6b-6f}^{(+-)} &=& \big( H_{3F}^{(+)} - G_{ 9B} \big)
                        - \big( H_{3B}^{(-)} - G_{10F} \big) \\
     H_{6b-6f}^{(-+)} &=& \big( H_{3F}^{(-)} - G_{10B} \big)
                        - \big( H_{3B}^{(+)} - G_{ 9F} \big) \\
   \end{eqnarray*}
  \end{minipage}
\hspace{12cm}
\end{appendix}
\end{widetext}

\end{document}